\documentclass[reprint,
superscriptaddress,
nofootinbib,
amsmath,amssymb,
aps,
pra,
floatfix]{revtex4-2}
\pdfoutput=1

\usepackage{graphicx}% Include figure files
\usepackage{dcolumn}% Align table columns on decimal point
\usepackage{bm}% bold math
\usepackage[colorlinks, citecolor={blue}, urlcolor={blue}, linkcolor={blue}]{hyperref}% add hypertext capabilities
\usepackage{multirow}
\usepackage{makecell}
\usepackage{braket}
\usepackage{array}
\newcolumntype{L}{>{\centering\arraybackslash}m{.12\textwidth}}
\newcolumntype{M}{>{\centering\arraybackslash}m{.2\textwidth}}
\newcolumntype{N}{>{\centering\arraybackslash}m{.08\textwidth}}
\newcolumntype{P}{>{\centering\arraybackslash}m{.17\textwidth}}
\newcolumntype{Q}{>{\arraybackslash}p{5cm}}

\begin{document}

\title{Comparison of Spontaneous Emission in Trapped Ion Multiqubit Gates at High Magnetic Fields}

\author{Allison L. Carter}
\email{allison.carter@nist.gov}
\affiliation{National Institute of Standards and Technology, Boulder, Colorado 80305, USA}
\author{Sean R. Muleady}
\affiliation{JILA, NIST and Department of Physics, University of Colorado, Boulder, Colorado 80309, USA}
\affiliation{Center for Theory of Quantum Matter, University of Colorado, Boulder, Colorado 80309, USA}
\author{Athreya Shankar}
\affiliation{Institute for Theoretical Physics, University of Innsbruck, 6020 Innsbruck, Austria}
\affiliation{Institute for Quantum Optics and Quantum Information of the Austrian Academy of Sciences, 6020 Innsbruck, Austria}
\author{Jennifer F. Lilieholm}
\affiliation{National Institute of Standards and Technology, Boulder, Colorado 80305, USA}
\affiliation{Department of Physics, University of Colorado, Boulder, Colorado 80309, USA}
\author{Bryce B. Bullock}
\affiliation{National Institute of Standards and Technology, Boulder, Colorado 80305, USA}
\affiliation{Department of Physics, University of Colorado, Boulder, Colorado 80309, USA}
\author{Matthew Affolter}
\affiliation{National Institute of Standards and Technology, Boulder, Colorado 80305, USA}
\author{Ana Maria Rey}
\affiliation{JILA, NIST and Department of Physics, University of Colorado, Boulder, Colorado 80309, USA}
\affiliation{Center for Theory of Quantum Matter, University of Colorado, Boulder, Colorado 80309, USA}
\author{John J. Bollinger}
\affiliation{National Institute of Standards and Technology, Boulder, Colorado 80305, USA}

\date{\today}% It is always \today, today,
             %  but any date may be explicitly specified

\begin{abstract}
 Penning traps have been used for performing quantum simulations and sensing with hundreds of ions and provide a promising route toward scaling up trapped ion quantum platforms because of the ability to trap and control hundreds or thousands of ions in two- and three-dimensional crystals. In both Penning traps and the more common radiofrequency Paul traps, lasers are often used to drive multiqubit entangling operations. A leading source of decoherence in these operations is off-resonant spontaneous emission. While many trapped ion quantum computers or simulators utilize clock qubits, other systems, especially those with high magnetic fields such as Penning traps, rely on Zeeman qubits, which require a more complex calculation of this decoherence. We therefore examine theoretically the impacts of spontaneous emission on quantum gates performed with trapped-ion ground state Zeeman qubits in a high magnetic field. In particular, we consider two types of gates\,---\,light-shift ($\hat{\sigma}_i^z\hat{\sigma}_j^z$) gates and M\o lmer-S\o rensen ($\hat{\sigma}_i^x\hat{\sigma}_j^x$) gates\,---\,obtained with laser beams directed approximately perpendicular to the magnetic field (the quantization axis) and compare the decoherence errors in each. Within each gate type, we also compare different operating points with regards to the detunings, polarizations, and required intensity of the laser beams used to drive the gates. We show that both gates can have similar performance at their optimal operating conditions at high magnetic fields and examine the experimental feasibility of various operating points. By examining the magnetic field dependence of each gate, we demonstrate that, when the $P$ state fine structure splitting is large compared to the Zeeman splittings, the theoretical performance of the M\o lmer-S\o rensen gate is significantly better than that of the light-shift gate. Additionally, for the light-shift gate, we make an approximate comparison between the fidelities that can be achieved at high fields with the fidelities of state-of-the-art two-qubit trapped ion quantum gates. We show that, with regard to spontaneous emission, the achievable infidelity with our current configuration is about an order of magnitude larger that of the best low-field gates, but we also discuss several alternative configurations with potential error rates that are comparable with those for state-of-the-art trapped ion gates. 
\end{abstract}

%\keywords{Suggested keywords}%Use showkeys class option if keyword
%display desired
\maketitle

%\tableofcontents

\section{Introduction}\label{sec:Introduction}
Trapped ions are a versatile quantum platform, with applications in computation \cite{bruzewicz2019}, simulation \cite{monroe2021}, and metrology \cite{brewer2019, gilmore2021, roos2006}. They are especially promising given their long qubit coherence times \cite{wang2021} and high gate fidelities \cite{ballance2016, gaebler2016, sutherland2020, tinkey2022, clark2021}. In linear radiofrequency (RF) traps, one-dimensional (1D) crystals of greater than 100 ions have been formed \cite{pagano2018}, and more than 50 ions have been employed in quantum simulation experiments \cite{zhang2017, kranzl2022, joshi2022}. In general, large ion crystals with more than one dimension are desirable for implementing quantum simulations that are more challenging to simulate with classical information and for increasing the sensitivity of quantum sensing. Moreover, simulations of certain many-body physics are more straightforward in systems with natively multidimensional structure \cite{richerme2016}. Penning ion traps, which use static electric and magnetic fields to confine ions, provide an opportunity to scale to larger two-dimensional (2D) and possibly three-dimensional (3D) ion crystals. Non-equilibrium quantum dynamics have been implemented with global couplings on single-plane crystals of greater than 200 ions in a Penning trap and benchmarked through the generation of entangled spin states \cite{bohnet2016, garttner2017}. The ease with which large 3D crystals can be formed and controlled in Penning traps provides motivation for investigating how quantum simulation and sensing can be extended to these much larger crystals \cite{itano1998}. We note that RF ion traps are also being investigated for quantum information processing with 2D and 3D crystals \cite{donofrio2021,qiao2022,mielenz2016,wu2021,duan2022,kiesenhofer2023}, but to date the number of ions used in experiments that generate entangled spin states has been modest ($< 20$). In this paper we examine different laser beam configurations, including laser beam polarizations and detunings from atomic resonances, for optimizing quantum simulation and sensing at the large magnetic fields suitable for Penning ion traps.

\begin{figure}[t]
    \centering
    \includegraphics[width=\columnwidth]{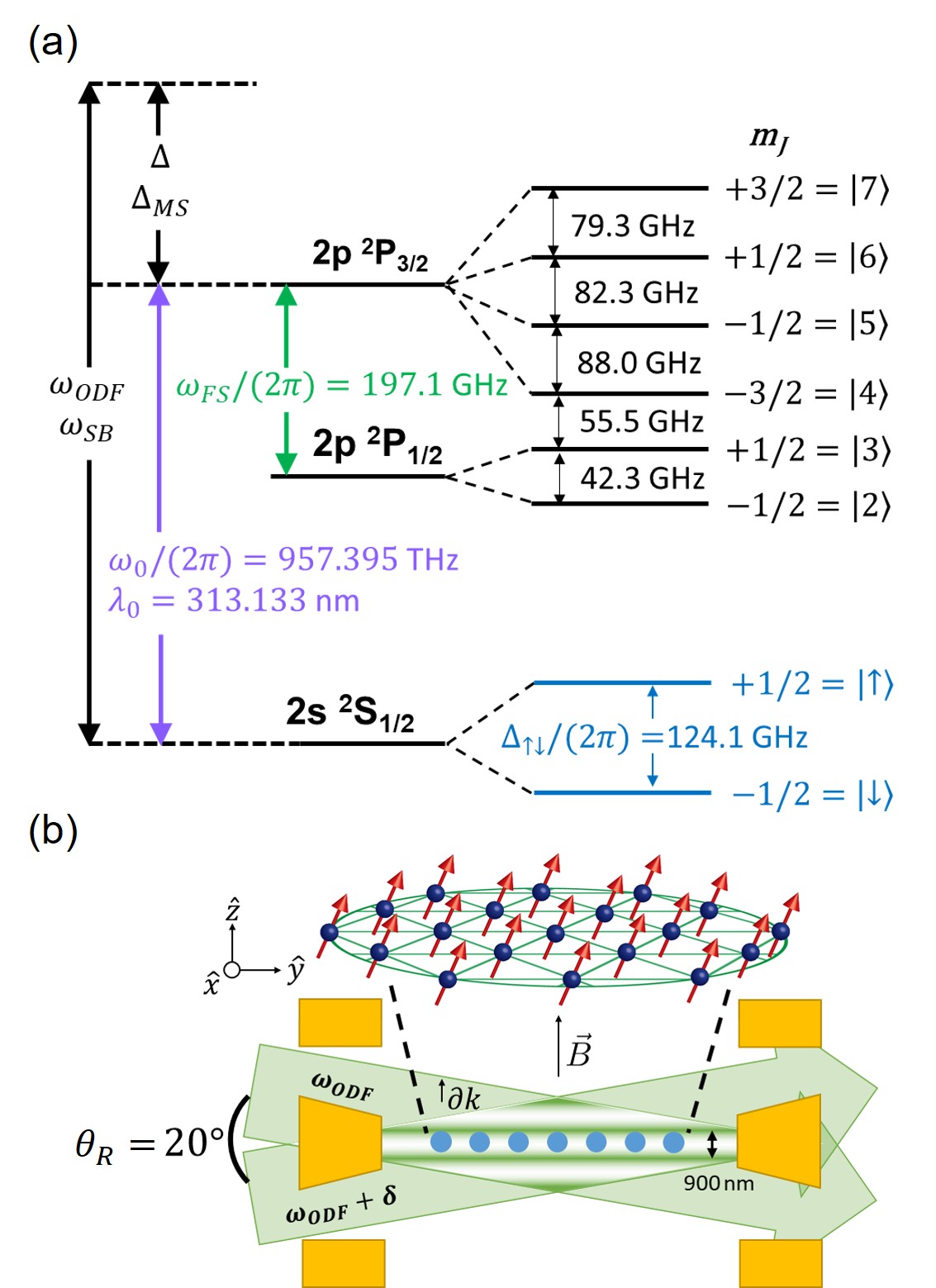}
    \caption{(a) Level structure of $^9$Be$^+$ at 4.46 T. The qubit is defined in the ground state $^2S_{1/2}$ manifold of Zeeman levels, shown here in blue. We define a magnetic field independent reference frequency $\omega_0$ as the frequency difference at zero magnetic field between the $S_{1/2}$ and $P_{3/2}$ states. The approximate frequencies and detunings for the optical dipole force (ODF) beams for the LS gate ($\omega_{ODF}$, $\Delta$) and for the sideband (SB) beams for the M\o lmer-S\o rensen (MS) gate ($\omega_{SB}$, $\Delta_{MS}$) are shown on the left. (b) Schematic of the laser beams driving the light shift gate and the ion crystal. The Penning trap electrodes are shown in yellow, ions are in blue, and laser beams are in light green. The beams enter the trap at angles of $\pm 10^\circ$ relative to the plane of the ion crystal and are detuned relative to each other by angular frequency $\delta$.}
    \label{fig:BeLevelsODFBeams}
\end{figure}

The large magnetic field required for trapping ions in a Penning trap ($\gtrsim 0.3$~T) generates large Zeeman splittings in the ground and excited states of the trapped ions and changes some of the considerations for optimizing the performance of a multiqubit entangling operation. Here we focus on Zeeman qubits, whose frequency splittings can range from tens of GHz to greater than 100~GHz at the high magnetic fields in Penning traps. We also focus our considerations on multiqubit entangling operations obtained by globally coupling the Zeeman qubits to a single motional mode or collection of motional modes with a spin-dependent optical dipole force from off-resonant laser beams. We show that tuning the frequency of the optical dipole force lasers between the large Zeeman splittings in the excited states produces some favorable features and discuss the trade-offs between tuning within versus outside the Zeeman manifold of excited states. The large laser beam waists required for global couplings with large ion crystals mean the available laser power can become a constraint, and we include this consideration in our discussion. Many possible sources of decoherence can affect trapped ion gates, such as motional heating, magnetic field fluctuations, and fluctuating parameters in the lasers used to drive operations. However, in two of the highest fidelity two-qubit entangling gates performed on ground state qubits reported to date, spontaneous emission was the largest source of error \cite{gaebler2016, ballance2016}. Spontaneous emission was also the leading source of decoherence in previous quantum simulation experiments with ion crystals in a Penning trap \cite{bohnet2016, garttner2017}. For this paper, we therefore focus on spontaneous emission as an error source. We carefully consider its impact for different operating conditions by examining figures of merit that compare the strength of the engineered spin-spin interaction to decoherence from off-resonant light scattering.

A common Hamiltonian for multiqubit gates in trapped ion quantum computers and simulators is the Ising Hamiltonian \cite{korenblit2012,jurcevic2014,garttner2017,figgatt2017,joshi2020}
\begin{equation}
    \hat{H}_{\textrm{Ising}} = \frac{\hbar}{\mathcal{N}}\sum_{i< j}J_{ij}\hat{\sigma}_i^\alpha\hat{\sigma}_j^\alpha,
    \label{eq:Ising_general}
\end{equation}
where $\hat{\sigma}_i^\alpha$ is one of the Pauli matrices $\hat{\sigma}^{x, y, z}$, $\mathcal{N}$ is the number of ions, and $J_{ij}$ is the coupling strength between ions $i$ and $j$. A leading type of entangling gate to implement this Hamiltonian is the geometric phase gate \cite{leibfried2003}. This gate employs spin-dependent forces to displace the ions in position-momentum phase space leading to a relative phase accumulation for different internal spin states. The size of this phase accumulation is proportional to the area enclosed by the loops driven in phase space \cite{leibfried2003,ge2019}. Two typical methods for implementing such a gate are the light-shift (LS) gate \cite{leibfried2003}, which relies on the AC Stark shift induced on the qubit states by the driving laser beam, and the M\o lmer-S\o rensen (MS) gate \cite{molmer1999,sorensen1999}. The effect of both gates on a system of $\mathcal{N}$ ions can be written in terms of the Ising Hamiltonian from Eq.~\ref{eq:Ising_general} where $\alpha \rightarrow z$ for the LS gate and $\alpha \rightarrow x$ or $\alpha \rightarrow y$ for the MS gate. In Secs.~\ref{sec:LSGateOptimization} and \ref{sec:MSGateHighFields}, we will discuss in more detail how $J_{ij}$ can be written in terms of relevant atomic and experimental parameters.

\begin{figure}[t]
    \centering
    \includegraphics[width=\columnwidth]{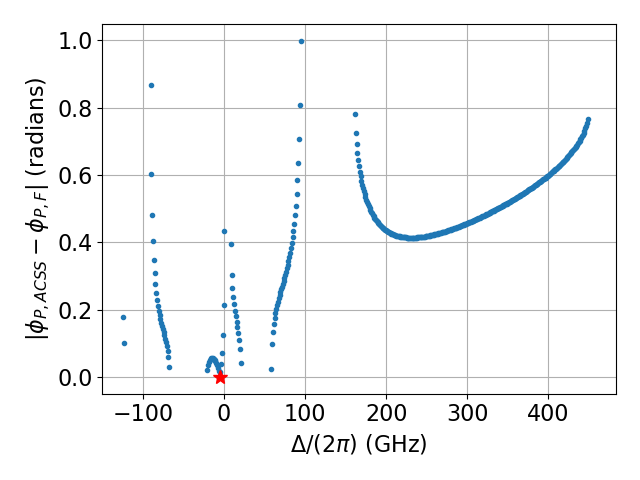}
    \caption{Plot of polarization differences between values of $\phi_{P, ACSS}$ and $\phi_{P,F}$ versus $\Delta$. Missing points have no polarization angle where $\Delta_{ACSS}=0$ or where $F_\uparrow = -F_\downarrow$. Note that at $\Delta/(2\pi)\approx-5$~GHz, there is a single point (indicated with a red star) with a difference of 0 rad between $\phi_{P, ACSS}$ and $\phi_{P, F}$.}
    \label{fig:ACSSNulls_Fup=Fdown}
\end{figure}

To consider a specific case for numerical calculations, we will use the operating parameters typical for the National Institute of Standards and Technology (NIST) Penning trap where single-plane crystals of $^9$Be$^+$ ions are employed for quantum simulations \cite{bohnet2016} and sensing \cite{gilmore2021}. The $^9$Be$^+$ atomic level structure at the 4.46~T magnetic field of the NIST Penning trap is shown in Fig.~\hyperref[fig:BeLevelsODFBeams]{\ref*{fig:BeLevelsODFBeams}(a)}. At these high magnetic fields, the Doppler cooling laser pumps the nuclear spin $(I=3/2)$ into a single projection state $\left(m_I=+3/2\right)$ \cite{itano1981}, resulting in a nuclear structure analogous to that of a nucleus with $I=0$. We define our qubit in the $^2S_{1/2}$ manifold of Zeeman sublevels, with $\ket{\downarrow} = \Ket{^2S_{1/2}, m_J=-1/2}$ and $\ket{\uparrow} = \Ket{^2S_{1/2}, m_J = +1/2}$. With a magnetic field $B$ of 4.46 T, the qubit splitting is approximately 124~GHz, which is much larger than that of a typical ground state qubit in a trapped ion in an RF Paul trap.

Previous work in the Penning trap at NIST has used an LS gate to drive multiqubit operations \cite{Britton2012,bohnet2016,garttner2017,gilmore2021,affolter2020}. The LS gate requires two laser beams that are detuned relative to each other by approximately the frequency of a motional mode of the ions, typically of order 1 MHz. More details are discussed in Sec.~\ref{sec:LSGateOptimization}. In some applications such as the motional sensing work discussed in Refs.~\onlinecite{gilmore2021} and \onlinecite{affolter2020}, the phase of the driven spin-dependent motion of the ions, which is sensitive to the phase difference between the two driving beams in the LS gate, is critical. Experimental constraints limit our ability to stabilize the relative phase of the two beams close to the ions, and as a result it is difficult to maintain robust phase stability as required for high-quality motional sensing. The nearly copropagating setup of the sideband beams for the MS gate, on the other hand, reduces the potential impact of differential path-length fluctuations between the red and blue sidebands \cite{haljan2005}. The MS gate would not only improve the phase stability in sensing work, but could also be used to drive spin-dependent motion in 3D crystals \cite{cui2022}. Employing 3D crystals could enable an increase in the number of ions employed in trapped-ion quantum simulation and sensing from several hundred ions to potentially $\sim 10^5$ ions \cite{itano1998}. 

In this paper, we begin in Sec.~\ref{sec:LSGateOptimization} by presenting a more detailed characterization of the previously reported NIST LS gate. We examine in detail the impact of spontaneous emission at 4.46~T for various operating points. In Sec.~\ref{sec:MSGateHighFields}, we examine the impact of spontaneous emission on the MS gate and show that, with regards to decoherence due to spontaneous emission, the MS gate is comparable to the LS gate for $^9$Be$^+$ in a magnetic field of 4.46~T. In Sec.~\ref{sec:VaryingB}, we examine how the performance of both the LS and MS gates depends on the magnetic field strength. In Sec.~\ref{sec:Fidelity}, we consider the fidelity limitations of a high magnetic field LS gate for a two-qubit system. In considering the fidelity of the two-qubit gate, we derive analytical expressions for the fidelity of the gate in terms of decoherence rates due to different types of off-resonant light scattering. Section~\ref{sec:OperatingPoints} gives a summary of various possible operating points for gates at high magnetic fields with a focus on the trade-offs between required laser power and improved performance as detunings from atomic resonances increase. In Sec.~\ref{sec:MgComparison}, we compare gates in $^9$Be$^+$ with those in heavier ions that have larger $^2P$ fine structure splittings. Finally in Sec.~\ref{sec:Conclusion}, we summarize our results.

\section{Optimizing a light-shift gate at a high magnetic field}\label{sec:LSGateOptimization}
\subsection{Background and current configuration}\label{sec:LSBackground}

In the current laser beam configuration employed at NIST (see Fig.~\hyperref[fig:BeLevelsODFBeams]{\ref*{fig:BeLevelsODFBeams}(b)}), a light-shift gate is implemented with two laser beams entering the trap at angles of $\pm \theta_R/2 = \pm 10^\circ$ with respect to the single-plane ion crystal. The small angles facilitate the alignment of the resulting 1D optical lattice wave vector $\partial\vec{k}$ to be parallel to the magnetic field and normal to the planar ion crystal. The electric fields for the upper and lower beams, respectively, are given by \cite{Britton2012}
\begin{align}
    \vec{\mathcal{E}}_u &= \mathcal{E}_0\hat{\epsilon}_u\cos\left[\vec{k}_u\cdot\vec{r} - \omega_{ODF}t\right],\nonumber\\
    \vec{\mathcal{E}}_l &= \mathcal{E}_0\hat{\epsilon}_l\cos\left[\vec{k}_l\cdot\vec{r}-\left(\omega_{ODF}+\delta\right)t\right],
\end{align}
where we have assumed the beams have equal intensity. $\hat{\epsilon}_{u(l)}$ is the polarization of the upper (lower) beam which, assuming linear polarization,\footnote{The assumption of linear polarization simplifies the calculation and provides the best performance for the configuration where the ODF laser beams are directed approximately perpendicular to the magnetic field as illustrated in Fig.~\hyperref[fig:BeLevelsODFBeams]{\ref*{fig:BeLevelsODFBeams}(b)}. We note that, independent of the ODF beam polarization, this perpendicular laser beam configuration dictates approximately equal coupling to $\sigma^+$ and $\sigma^-$ transitions.} can be parameterized in terms of angles $\phi_{P_{u(l)}}$ as $\hat{\epsilon}_{u(l)} = \cos\left(\phi_{P_{u(l)}}\right)\hat{z} + \sin\left(\phi_{P_{u(l)}}\right)\hat{x}$. The frequency difference between the two beams is parameterized by $\delta$. We define $\hbar\omega_0$ as the energy difference between the $^2S_{1/2}$ and $^2P_{3/2}$ levels at zero magnetic field. The detuning of the laser beams from atomic resonances $\Delta$ is defined in relation to $\omega_0$ as $\Delta = \omega_{ODF} - \omega_0$ as shown in Fig.~\hyperref[fig:BeLevelsODFBeams] {\ref*{fig:BeLevelsODFBeams}(a)}. 

Each beam produces an AC Stark shift (ACSS) on each qubit state,
\begin{eqnarray}
\Delta_{ACSS,\, \uparrow, i} &= A_\uparrow \cos^2\left(\phi_{P_i}\right) + B_\uparrow\sin^2\left(\phi_{P_i}\right),\nonumber\\
\Delta_{ACSS,\, \downarrow, i} &= A_\downarrow \cos^2\left(\phi_{P_i}\right) + B_\downarrow\sin^2\left(\phi_{P_i}\right),
\label{eq:ACSS_Individual}
\end{eqnarray}
where the index $i$ indicates the upper or lower beam. $A_\uparrow$ and $A_\downarrow$ are the Stark shifts on $\ket{\uparrow}$ and $\ket{\downarrow}$, respectively, from a purely vertically polarized beam ($\pi$ polarization) and $B_\uparrow$ and $B_\downarrow$ are the corresponding Stark shifts from a purely horizontally polarized beam, which contains an equal superposition of $\sigma^+$ and $\sigma^-$ light. We are making a small angle approximation here such that vertical polarization corresponds exactly to $\pi$ polarization, which would be exact only for $\theta_R/2 = 0$.

We assume the frequency offset $\delta$ between the laser beams will be small compared to the detuning of the laser beams from any atomic transition. With this assumption, the coefficients $A_\downarrow, A_\uparrow, B_\downarrow$, and $B_\uparrow$ are essentially independent of $\delta$ and given by
\begin{eqnarray}
A_\downarrow &= \left(\frac{g_0}{\mu}\right)^2\sum_j \frac{\left|\Braket{j|\vec{d}\cdot\hat{z}|\downarrow}\right|^2}{\Delta + \left(\omega_0-\omega_j\right)},\nonumber\\
B_\downarrow &= \left(\frac{g_0}{\mu}\right)^2\sum_j\frac{\left|\Braket{j |\vec{d}\cdot\hat{x}|\downarrow}\right|^2}{\Delta + \left(\omega_0 - \omega_j\right)},\nonumber\\
A_\uparrow &= \left(\frac{g_0}{\mu}\right)^2\sum_j\frac{\left|\Braket{ j|\vec{d}\cdot\hat{z}|\uparrow}\right|^2}{\Delta + \Delta_{\uparrow\downarrow} +\left(\omega_0-\omega_j\right)},\nonumber\\
B_\uparrow &= \left(\frac{g_0}{\mu}\right)^2\sum_j\frac{\left|\Braket{ j|\vec{d}\cdot\hat{x}|\uparrow}\right|^2}{\Delta + \Delta_{\uparrow\downarrow} + \left(\omega_0-\omega_j\right)},
\label{eq:ACSS_coeffs}
\end{eqnarray}
where $g_0 = \mathcal{E}_0\mu/(2\hbar)$ is the single-photon Rabi frequency, $\mu$ is the largest matrix element $\Braket{\uparrow|\vec{d}\cdot\hat{\epsilon}_{1}|P_{3/2}, m_J = +3/2}$, $\hat{\epsilon}_1$ is the polarization vector for $\sigma^+$ polarized light, and $\vec{d}$ is the dipole operator. Additionally, the sums are taken over all excited states, $\Delta_{\uparrow\downarrow}$ is the qubit splitting, and $\omega_j$ is the frequency of the transition $\ket{\downarrow} \leftrightarrow \ket{j}$. 

We are particularly interested in the differential ACSS, $\Delta_{ACSS} \equiv \Delta_{ACSS,\, \uparrow} - \Delta_{ACSS,\,\downarrow}$. If $\Delta_{ACSS} = 0$, the sensitivity of any operation to fluctuations in the laser intensity will be dramatically reduced resulting in more robust operations. The differential ACSS for each beam is given by
\begin{equation}
    \Delta_{ACSS_i} = \left(A_\uparrow-A_\downarrow\right)\cos^2\left(\phi_{P_i}\right) + \left(B_\uparrow-B_\downarrow\right)\sin^2\left(\phi_{P_i}\right).
    \label{eq:ACSS_diff}
\end{equation}
Below, we show that we can null the differential ACSS for each beam individually while maintaining a spin-dependent force. This condition on the individual beams results in further robustness to laser intensity fluctuations.

From this equation combined with Eq.~\ref{eq:ACSS_coeffs}, it is clear that the differential ACSS depends on both the detuning $\Delta$ and the polarization angle $\phi_{P_i}$. These equations can then be solved numerically to find values of $\Delta$ and $\phi_{P_i}$ where the differential ACSS will be nulled. Such values can be found for detunings where $A_\uparrow-A_\downarrow$ and $B_\uparrow-B_\downarrow$ have opposite signs \cite{Britton2012}. If, for a given value of $\Delta$, the differential ACSS is nulled for a polarization angle $\phi_{P, ACSS}$, the angle $-\phi_{P, ACSS}$ will also give $\Delta_{ACSS_i}=0$.

If we assume $\phi_P \equiv \phi_{P_u}= -\phi_{P_l}$, applying both beams results in a spin-dependent force $F_{\uparrow(\downarrow)} = F_{0_{\uparrow(\downarrow)}}\sin\left(\partial k\cdot z-\delta t\right)$ \cite{Britton2012}, where
\begin{equation}
    F_{0_{\uparrow(\downarrow)}} = -2\hbar \partial k \left[A_{\uparrow(\downarrow)}\cos^2\left(\phi_P\right) - B_{\uparrow(\downarrow)}\sin^2\left(\phi_P\right)\right],
    \label{eq:F_0updown}
\end{equation}
and $\partial k= \left|\vec{k}_u-\vec{k}_d\right|$ is the wave vector difference between the upper and lower beams.

The choice $\phi_{P_u} = -\phi_{P_l}$ allows us to null the differential ACSS for each beam but still have a nonzero spin-dependent force. The condition $F_\uparrow = -F_\downarrow$ also increases how robust the gate is to laser intensity fluctuations \cite{Britton2012}, so we try to satisfy this condition as well. We define the polarization angle that nulls the differential ACSS as $\phi_{P, ACSS}$ and the polarization angle that results in $F_\uparrow = -F_\downarrow$ as $\phi_{P, F}$. For most detunings, $\phi_{P, ACSS}$ does not equal $\phi_{P, F}$. In Fig.~\ref{fig:ACSSNulls_Fup=Fdown}, we plot the difference between $\phi_{P, ACSS}$ and $\phi_{P,F}$ versus the laser detuning $\Delta$. There is a single point where one polarization angle satisfies both conditions. It occurs at $\Delta/(2\pi)=-5.29$~GHz with $\phi_P = \phi_{P, ACSS} = \phi_{P,F} =\pm 65.25^\circ$ at 4.46 T. This is the current operating point for the NIST LS gate \cite{Britton2012,bohnet2016}.

\subsection{Spontaneous emission in the light-shift gate}\label{sec:LSSpontaneousEmission}

In previous quantum simulation results employing the configuration and operating point discussed in the previous section, spontaneous emission was the leading source of decoherence \cite{bohnet2016}. Here we present a thorough analysis of this decoherence. Our aim is, first, to characterize better our current operating point and explore if there are better operating points for the LS gate, and, second, to compare the performance of the LS gate with that of the M\o lmer-S\o rensen gate discussed in Sec.~\ref{sec:MSGateHighFields}.

To calculate the rate of decoherence due to spontaneous emission, we follow the formalism outlined in Ref.~\onlinecite{Uys2010}, which considered single-qubit decoherence from a single off-resonant laser beam. There are two relevant categories of spontaneous emission: Raman (inelastic or state-changing) and Rayleigh (elastic or non-state-changing) scattering. In many other results for trapped ions, such as Refs.~\onlinecite{gaebler2016} and \onlinecite{ozeri2007}, decoherence due to Rayleigh scattering is either assumed or shown to be negligible. This is not the case for many potential configurations for entangling gates in high magnetic fields. 

To summarize the results from Ref.~\onlinecite{Uys2010}, the rate of Raman scattering from a starting state $\ket{i}$ to a final state $\ket{j}$ can be calculated using the Kramers-Heisenberg formula,
\begin{equation}
    \Gamma_{ij, 1}=g_0^2\gamma\sum_\lambda\left(\sum_J A_{J,\lambda}^{i\rightarrow j}\right)^2,
    \label{eq:Gamma_ij_singlebeam}
\end{equation}
where $\gamma$ is the rate of spontaneous decay from the relevant excited state, $\lambda$ indicates the polarization of the photon, and $J \in \left\{\frac{1}{2},\frac{3}{2}\right\}$ for the corresponding $P$ fine structure levels. The subscript 1 indicates that this is the rate for a single laser beam. The absence of such a subscript in later equations will indicate we are considering the total rate of decoherence due to both laser beams. The single-beam Raman scattering decoherence rate is defined as,
\begin{equation}
    \Gamma_{r,1} = \Gamma_{\uparrow\downarrow, 1} + \Gamma_{\downarrow\uparrow, 1}.
    \label{eq:Gamma_r_ZZ_singlebeam}
\end{equation}
For our calculations, the polarization $\lambda$ is 0 for $\pi$ transitions and $\pm 1$ for $\sigma^\pm$ transitions. We use $\hat{\epsilon}_{0,\pm 1}$ respectively to indicate the polarization vectors. The relevant scattering amplitudes $A_{J,0}^{i\rightarrow j}$ and $A_{J,\pm 1}^{i\rightarrow j}$ are then
\begin{equation}
    A_{J,\lambda}^{i\rightarrow j} = \frac{b_\lambda\Braket{j|\vec{d}\cdot\hat{\epsilon}_{\lambda + i-j}^*|J,i +\lambda}\Braket{J, i + \lambda |\vec{d}\cdot\hat{\epsilon}_\lambda|i}}{\Delta_{iJ\lambda}\mu^2}, 
    \label{eq:ScatteringAmplitudes}
\end{equation}
where $i, j \in \left\{-\frac{1}{2},+\frac{1}{2}\right\}$, $b_0 = \cos\left(\phi_P\right)$, and $b_{\pm 1} = \frac{1}{\sqrt{2}}\sin\left(\phi_P\right)$. States that are given by $\ket{i}$ or $\ket{j}$ are qubit states and described only by their $m_J$ values, while excited states are described by $\Ket{J, m_J}$. $\Delta_{iJ\lambda}$ can be written as $\Delta_{iJ\lambda} = \Delta + \left(\omega_0 - \omega_{J,i+\lambda}\right)$ if $i = -\frac{1}{2}$ or $\Delta_{iJ\lambda} = \Delta + \Delta_{\uparrow\downarrow}  + \left(\omega_0 - \omega_{J, i+\lambda}\right)$ if $i = +\frac{1}{2}$. The frequency $\omega_{J, i +\lambda}$ is defined such that $\hbar\omega_{J, i+\lambda}$ is the energy difference of $\ket{J, i+\lambda}$ from the lowest energy level $\Ket{^2S_{1/2}, m_J = -1/2}$. We note that for very large detunings (hundreds of THz) there are additional corrections required for Eq.~\ref{eq:Gamma_ij_singlebeam} as discussed in Ref.~\onlinecite{moore2022}. However, for the detunings we consider in this paper, these additional factors should remain small.

Qubit decoherence due to Rayleigh scattering is calculated from the difference in the elastic scattering amplitudes for the two qubit levels. In particular, the decoherence rate for Rayleigh scattering is given by
\begin{equation}
    \Gamma_{el, 1} = g_0^2\gamma\sum_\lambda\left(\sum_J A_{J,\, \lambda}^{\downarrow \rightarrow \downarrow}-\sum_{J'}A_{J',\, \lambda}^{\uparrow\rightarrow\uparrow}\right)^2.
    \label{eq:Gamma_el_ZZ_singlebeam}
\end{equation}
If the scattering amplitudes for the two qubit states are nearly the same in both magnitude and sign, the decoherence rate will be very small. On the other hand, if the scattering amplitudes are similar in magnitude but opposite in sign, then the scattering amplitudes constructively interfere, and the decoherence rate is approximately twice that expected from the sum of the elastic scattering rates of the two-qubit states.

Reference~\onlinecite{Uys2010} shows that the total single-qubit, single laser beam decoherence rate, defined as the decay rate of the off-diagonal elements of the density matrix, is
\begin{equation}
    \Gamma_1 \equiv \frac{1}{2}\left(\Gamma_{r,1} + \Gamma_{el, 1}\right).
    \label{eq:Gamma_singleBeam_LS}
\end{equation}
Since the LS gate requires two laser beams at approximately the same frequency, however, we will define $\Gamma_{\uparrow\downarrow}$, $\Gamma_{\downarrow\uparrow}$, and $\Gamma_{el}$ as twice what would be obtained from a single beam. The expressions from Eqs.~\ref{eq:Gamma_ij_singlebeam}, \ref{eq:Gamma_r_ZZ_singlebeam}, \ref{eq:Gamma_el_ZZ_singlebeam}, and \ref{eq:Gamma_singleBeam_LS} are modified such that
\begin{align}
    \Gamma_{ij} &= 2g_0^2\gamma\sum_\lambda\left(\sum_J A_{J,\lambda}^{i\rightarrow j}\right)^2, \label{eq:Gamma_ij} \\
    \Gamma_r &= 2\left(\Gamma_{\uparrow\downarrow, 1} + \Gamma_{\downarrow\uparrow, 1}\right) = \Gamma_{\uparrow\downarrow} + \Gamma_{\downarrow\uparrow},\label{eq:Gamma_r_LS}\\ 
    \Gamma_{el} &= 2g_0^2\gamma\sum_\lambda\left(\sum_J A_{J,\lambda}^{\downarrow\rightarrow\downarrow} - \sum_{J'}A_{J',\lambda}^{\uparrow\rightarrow\uparrow}\right)^2, \label{eq:Gamma_el}\\
    \Gamma &\equiv \frac{1}{2}\left(\Gamma_r + \Gamma_{el}\right)=\left(\Gamma_{r,1} + \Gamma_{el, 1}\right). 
    \label{eq:Gamma_LS}
\end{align}

The inclusion of a coherent interaction with the analysis of Ref.~\onlinecite{Uys2010} can modify the effective single-qubit decoherence rate (see Appendix~\ref{sec:AppA}). For the LS gate the coherent interaction engineered by the spin-dependent optical dipole forces is the $\hat{\sigma}_i^z\hat{\sigma}_j^z$ Ising interaction given by Eq.~\ref{eq:Ising_general} with $\alpha \rightarrow z$. In the mean field approximation, this Ising interaction reduces to a $\hat{\sigma}_z$ interaction where each spin $\hat{\sigma}_i^z$ interacts with an effective magnetic field,
\begin{equation}
    \bar{B}_i =  \frac{1}{\mathcal{N}}\sum_{j\neq i}^{\mathcal{N}-1} J_{ij} \left\langle \hat{\sigma}_j^z \right\rangle,
    \label{eq:Beff_MF}
\end{equation}
determined by the mean value of the remaining spins \cite{Britton2012}. We show in Appendix~\ref{sec:AppA} that including a $\hat{\sigma}^z$ interaction in the master equation that describes decoherence from off-resonant light scatter does not change the effective single-spin decoherence rate from that given by Eq.~\ref{eq:Gamma_LS}. However, the $\hat{\sigma}^x$ interaction for the MS gate does in fact modify that decay rate, as discussed further in Sec.~\ref{sec:MSSpontaneousEmission}.

\subsection{Figure of merit for comparing spontaneous emission and interaction strength}\label{sec:FOM}
\begin{figure*}[t]
    \centering
    \includegraphics[width=\textwidth]{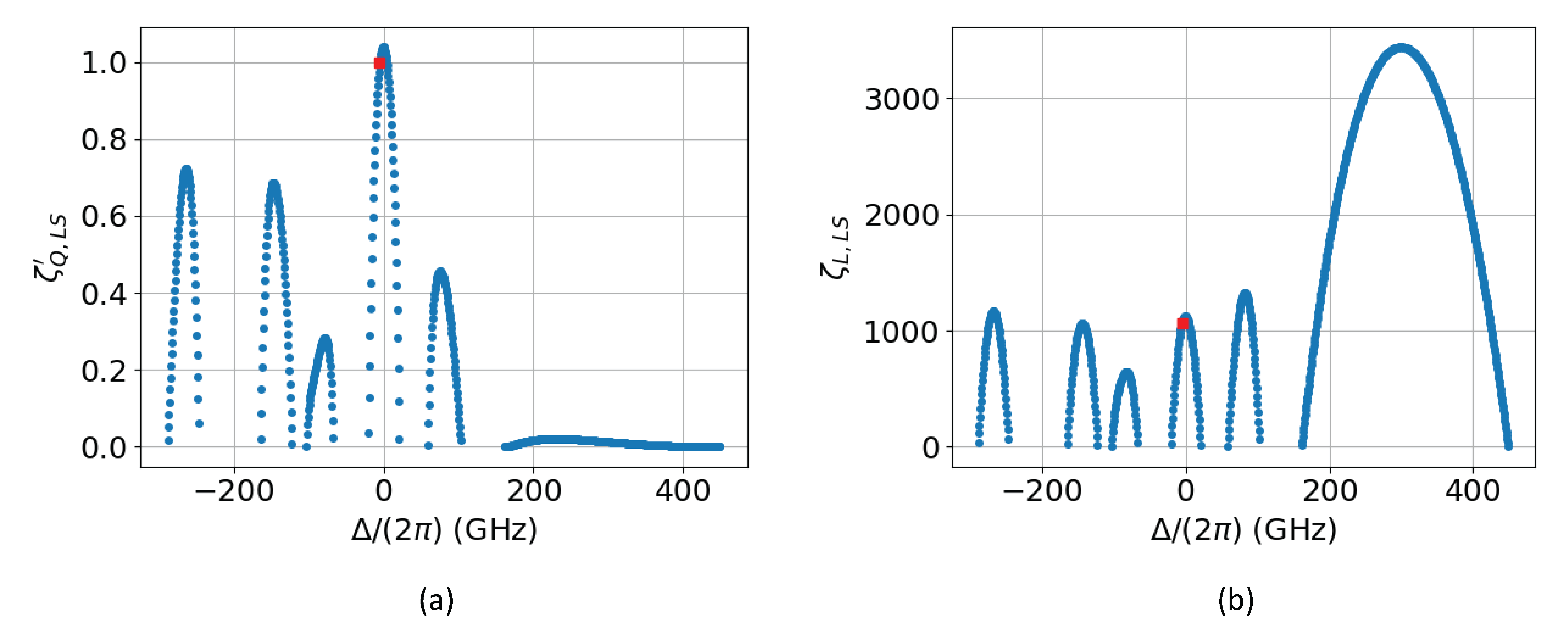}
    \caption{Plots of the figures of merit for the LS gate versus detuning with the differential ACSS of each laser beam nulled. The red square point indicates the current NIST operating point in both (a) and (b). For each point, the polarization angle $\phi_{P, ACSS}$ that nulls the ACSS is computed and then that polarization is used to compute $\zeta^\prime_{Q, LS}\left(\Delta,\phi_{P, ACSS}\right)$ and $\zeta_{L,LS}\left(\Delta,\phi_{P, ACSS}\right)$. (a) Plot of the normalized quadratic figure of merit $\zeta^\prime_{Q, LS}$ versus the detuning $\Delta/(2\pi)$ assuming $\Delta_{ACSS}=0$. No points are plotted for values of $\Delta/(2\pi)$ that have no corresponding value of $\phi_{P, ACSS}$. (b) Plot of $\zeta_{L, LS}$ versus $\Delta/(2\pi)$ with $\Delta_{ACSS}=0$. In (a) we assume the laser intensity is held fixed for all $\Delta$. However, $zeta_{L, LS}$ is independent of laser intensity, so the plot in (b) will remain the same, even if different laser intensities are used for different values of $\Delta$.}
    \label{fig:FOMZZDeltaACSS=0}
\end{figure*}
The decoherence rate given by Eq.~\ref{eq:Gamma_LS} and the spin-dependent force  will both approach zero for large detunings, and the trade-off between the interaction strength and the decoherence rate is not necessarily obvious. We construct metrics that will compare the strength of the engineered interaction between ion qubits to the rate of decoherence due to off-resonant light scatter. 

The pairwise interaction strength between ion qubits $i$ and $j$ is given by $J_{ij}$ in Eq.~\ref{eq:Ising_general}. More specifically, $J_{ij}$ can be written as \cite{kim2009, Britton2012}
\begin{equation}
    J_{ij} = \frac{F_0^2 \mathcal{N}}{2\hbar m}\sum_{n=1}^\mathcal{N} \frac{b_{i, n} b_{j, n}}{\delta^2 - \omega_n^2},
    \label{eq:Jij_LS}
\end{equation}
where $b_{i, n}$ is the component of the normalized eigenvector for mode $n$ and ion $i$. $\omega_n$ is the corresponding motional angular frequency, $m$ is the mass of the ions, and we have defined $F_0$ as 
\begin{equation}
    F_0 \equiv \frac{1}{2}\left(F_{0_\uparrow}-F_{0_\downarrow}\right).
    \label{eq:F0}
\end{equation}
For simplicity, we assume the modes to which we are coupling in Eq.~\ref{eq:Jij_LS} are transverse modes and $\omega_1 = \omega_z$ is the highest frequency mode, the center-of-mass (COM) mode.

We consider two regimes for multiqubit operations. The first regime is when the spin-dependent force and COM mode frequency difference $\delta_z \equiv \delta - \omega_z$ is comparable or large compared to the bandwidth of the transverse modes, in which case the coupling of the spin-dependent force to all modes is roughly equal. This regime is often used for quantum simulation \cite{joshi2020, tan2021, blatt2012, lin2011}. Here $\delta$ can be set arbitrarily as long as it remains far from any mode \cite{monroe2021}. In this ``quantum simulation" regime, the spin-dependent force drives many small rapidly accumulating phase space loops of each mode. The amplitude of the loops must be small compared to the size of the ground state wave function to avoid spin-motion entanglement and realize an effective spin model at all times. In this regime, the accumulated geometric phase for each mode grows linearly with time $t$ for time scales that are long compared to $2\pi/\delta_z$. It is reasonable for this case to hold $\delta_z$ and $\omega_z$ constant and define a figure of merit,
\begin{equation}
    \zeta_{Q, LS}\left(\Delta,\phi_P\right)\equiv \frac{\left[F_0/(\hbar \partial k)\right]^2}{\Gamma},
    \label{eq:FOM_ZZ}
\end{equation}
such that $\zeta_{Q, LS}\propto J_{ij}/\Gamma$. Here $\Gamma$ is the single-qubit decoherence rate given by Eq.~\ref{eq:Gamma_LS}. We will refer to $\zeta_{Q, LS}$ as the quadratic figure of merit. 

We note that $\zeta_{Q, LS}$ has units of s$^{-1}$ and depends linearly on the intensity of the laser beams used, since $F_0^2 \propto g_0^4 \propto \mathcal{E}^4$ while $\Gamma\propto g_0^2\propto \mathcal{E}^2$. For the purpose of comparing relative values of $zeta_{Q, LS}$ for different detunings and for the LS and MS (Sec.~\ref{sec:MSGateHighFields}) gates, we normalize the quadratic figures of merit by dividing by the value of $\zeta_{Q,LS}$ at the current NIST operating point $\zeta_{Q,0}$ (see Fig.~\hyperref[fig:FOMZZDeltaACSS=0]{\ref*{fig:FOMZZDeltaACSS=0}(a)}) assuming equal laser intensities for all points. This normalization results in a unitless parameter $\zeta^\prime_{Q,LS}$. For a laser intensity $I$ in units of W/m$^2$, $\zeta_{Q, 0}/I\sim 8400 \textrm{m}^2/(\textrm{Ws})$.

Because the spin-dependent force $F_0$ decreases with increasing detuning $\Delta$, large detunings permit higher laser intensities before the onset of spin-motion entanglement. The assumption of constant laser intensity as $\Delta$ is varied therefore impacts the value of the quadratic figure of merit $\zeta_{Q,LS}$ that we calculate at large detunings. This means $\zeta_{Q,LS}$ is appropriate to use only when laser power is a constraint and assumed to be constant over the bandwidth of detunings $\Delta$ being considered. Because constraining the laser intensity limits the potential performance of operations at large detunings, we will restrict our consideration of this figure of merit to detunings within approximately $\pm 200$~GHz of transitions to the $P$ state manifolds.
\begin{figure*}[t]
    \centering
    \includegraphics[width=\textwidth]{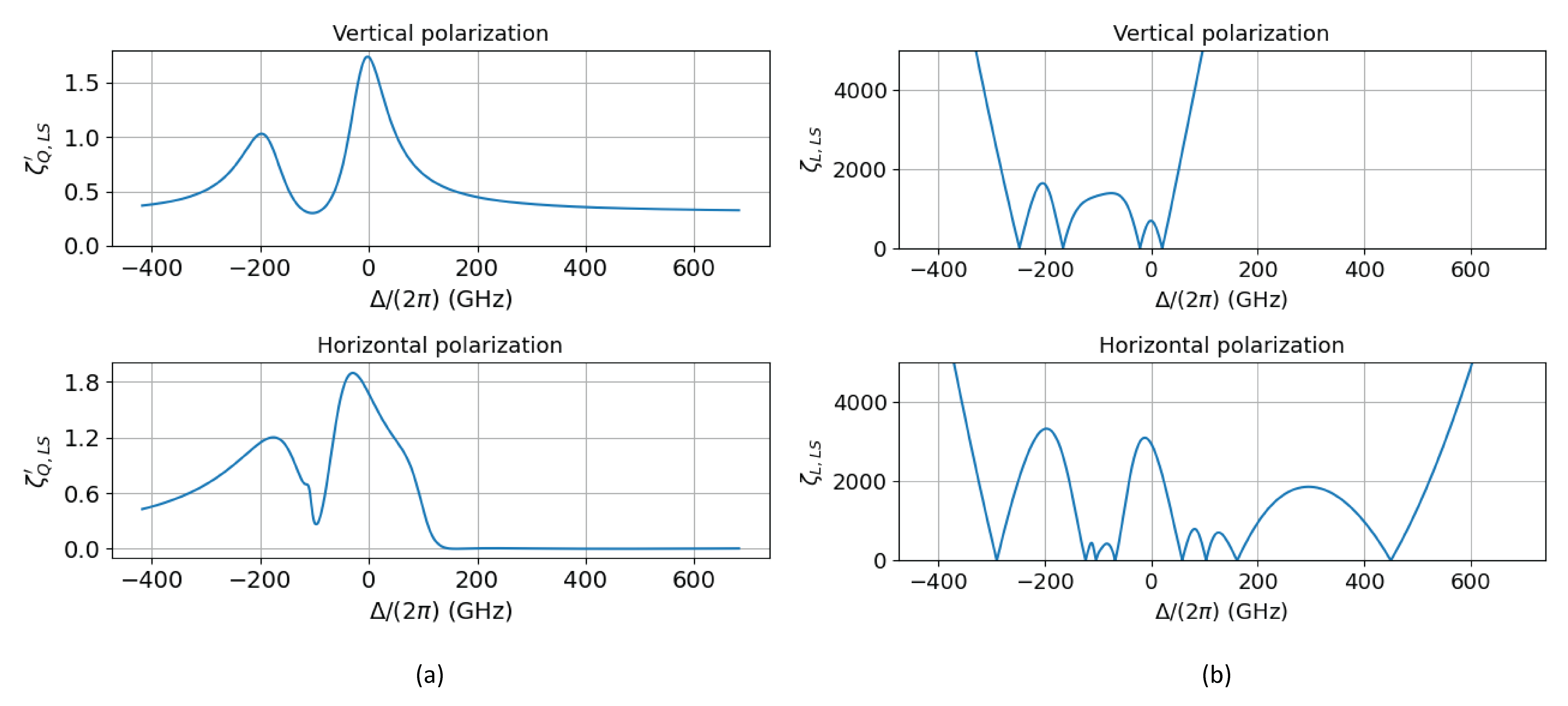}
    \caption{(a) Plot of $\zeta^\prime_{Q, LS}$ versus $\Delta/(2\pi)$ for vertical (top) and horizontal (bottom) polarizations. This figure of merit approaches constant values for large detunings, but is not useful in this regime because of the assumption of constant laser power. (b) Plot of $\zeta_{L,LS}$ versus $\Delta/(2\pi)$ for vertical (top) and horizontal (bottom) polarizations. Note that $\zeta_{L,LS}$ diverges rapidly for detunings outside the $P$ manifolds, indicating that spontaneous emission error will be rapidly suppressed by using large detunings when operating in the gate regime. Large detunings do require much higher power, however. The trade off between the challenges of the power requirements and the benefits of lower spontaneous emission are discussed further in Sec.~\ref{sec:OperatingPoints}. For horizontal polarization, the six points with the furthest negative detunings where $\zeta_{L,LS} =0$ correspond to resonances where $\Gamma \sim \gamma$. The points near $\Delta/(2\pi) \sim 170$ and 450~GHz with $\zeta_{L,LS}=0$ correspond to points where $\Delta_{ACSS}=0$ so the spin-dependent force is 0.}
    \label{fig:FOMZZVandH}
\end{figure*}
The second regime that we will consider is detuning close to a single motional mode such that the mode dominates the interactions. This is the regime relevant for experimental work done at NIST, where the spins are coupled mainly to the COM mode. If we assume that we couple primarily to the COM mode, $J_{ij}$ is constant for any pair of ions, and we have \cite{bohnet2016}
\begin{equation}
    J_{ij} \rightarrow J \approx \frac{F_0^2}{4\hbar m\omega_z \delta_z}. \label{eq:JCOM}
\end{equation}

This regime requires a more careful consideration. Here, the spin-dependent force drives large displacements in phase space, and if the ion crystal does not return to the original position in phase space at the end of the gate, there will be residual undesired spin-motion entanglement \cite{kim2009}. In this ``gate" regime one typically targets generating an entangled spin state obtained through the accumulation of a given amount of geometric phase $\Phi$ \cite{ge2019, ge2019b}. Optimal performance is achieved by executing one or a small number of complete phase space loops. In this case, one can show that
\begin{equation}
    \Phi \propto \left(F_0 \tau_g\right)^2,
    \label{eq:GeoPhase}
\end{equation}
where $\tau_g = 2\pi/\delta_z$ is the duration of the gate. Since, in this regime, geometric phase is accumulated quadratically with the gate duration, $\tau_g\propto \sqrt{\Phi}/F_0$, the anticipated error in the gate will be approximately $\Gamma\tau_g\sim \Gamma/F_0$. For the gate regime, therefore, we define a linear figure of merit,
\begin{equation}
    \zeta_{L,LS}\left(\Delta, \phi_P\right) \equiv \frac{F_0/(\hbar\partial k)}{\Gamma}. \label{eq:LS_LinFOM}
\end{equation}
We note that $\zeta_{L,LS}$ is dimensionless and independent of the laser intensity. 

In Fig.~\ref{fig:FOMZZDeltaACSS=0}, we plot the two figures of merit, $\zeta^\prime_{Q, LS}$ and $\zeta_{L, LS}$, versus the detuning $\Delta/(2\pi)$ for values of $\Delta$ where the differential ACSS is 0. The local maxima in the figures of merit correspond to laser detunings that lie between different Zeeman levels of the $P$ state manifold. Where the figure of merit approaches zero corresponds to resonances between one of the qubit levels and one of the Zeeman levels of the $P$ state manifold. The points in these plots are computed by finding the values of $\phi_P$ such that $\Delta_{ACSS}=0$ ($\phi_{P, ACSS}$) and then substituting that value along with $\Delta$ into Eqs.~\ref{eq:F_0updown}, \ref{eq:Gamma_r_LS}, and \ref{eq:Gamma_el} to compute $\zeta_{Q, LS}\left(\Delta, \phi_{P, ACSS}\right)$ and $\zeta_{L,LS}\left(\Delta, \phi_{P, ACSS}\right)$. We then divide all values of $\zeta_{Q, LS}$ by $\zeta_{Q,0}$ to obtain $\zeta^\prime_{Q, LS}$. As mentioned in Sec.~\ref{sec:LSBackground}, most values of $\Delta$ have no value $\phi_P$ that satisfies $\Delta_{ACSS}=0$, so we do not plot points for these detunings in Fig.~\ref{fig:FOMZZDeltaACSS=0}.

Furthermore, the figure of merit for the operating point we found in Sec.~\ref{sec:LSBackground} where $\Delta_{ACSS}=0$ and $F_\uparrow=-F_\downarrow$ is shown in red in both plots. Not only is this a favorable point to operate for robustness to laser intensity fluctuations, but for detuning between the excited state Zeeman levels, it also has a high value of $\zeta_{L, LS}$, indicating that our error rate due to spontaneous emission will be low compared to most other detunings.

There are some notable differences between the two figures of merit. In general the linear figure of merit increases at large detunings because $F_0\propto1/\Delta^2$ and $\Gamma\propto 1/\Delta^4$ (see Sec.~\ref{sec:VaryingB} and Appendix~\ref{appendix:BomegaFS_dependence}) at the expense of decreasing gate speed for fixed laser intensity. Another significant difference is the large peak in the value of $\zeta_{L, LS}$ with nulls of the ACSS over a bandwidth of $\sim 100$~GHz centered around $\Delta/(2\pi)\sim 300$~GHz. The presence of this peak provides an opportunity for both nulling the ACSS and suppressing errors due to spontaneous emission compared to the current NIST operating point when operating in the gate regime. However, we will show in Sec.~\ref{sec:OperatingPoints} that this point requires prohibitively high laser power.

Alternatively, we can relax the constraint on the ACSS and consider other possible configurations. If these configurations improve the appropriate figure of merit and if the laser power can be sufficiently stabilized, it may be worthwhile to consider operating with $\Delta_{ACSS}\neq0$. As two particular examples, we consider the cases of purely vertical or horizontal polarization in each beam and plot the results for both figures of merit in Fig.~\ref{fig:FOMZZVandH}. For the linear figures of merit for both horizontal and vertical polarization (see Fig.~\hyperref[fig:FOMZZVandH]{\ref*{fig:FOMZZVandH}(b)}) and the quadratic figure of merit with vertical polarization (top plot of Fig.~\hyperref[fig:FOMZZVandH]{\ref*{fig:FOMZZVandH}(a)}), the peaks are approximately centered between atomic resonances. The more complicated shape for the quadratic figure of merit with horizontal polarization results from transitions driven by both $\sigma^+$ and $\sigma^-$ polarized light. The plots in Fig.~\ref{fig:FOMZZVandH} also show that the maximum value of $\zeta_{Q, LS}$ is larger than $\zeta_{Q,0}$ by almost a factor of two, while $\zeta_{L,LS}$ grows rapidly for large detunings. For detuning $\delta$ close to a single motional mode, where using $\zeta_{L,LS}$ is appropriate, it would then be advantageous to have as large of a detuning $\Delta$ as possible if laser power is not a constraint and the laser intensity, and therefore the ACSS, can be sufficiently stabilized.

The results of this section assume $^9$Be$^+$ in a 4.46~T magnetic field. In Sec.~\ref{sec:VaryingB} we investigate how the figures of merit for the LS gate configuration discussed in Sec.~\ref{sec:LSBackground} depend on the magnetic field. In Sec.~\ref{sec:MgComparison} we indicate how the results discussed here are modified for ions with larger $P$ state fine structure splittings.

\section{High-field M\o lmer-S\o rensen gate}\label{sec:MSGateHighFields}

\begin{figure}[t]
    \centering
    \includegraphics[width=\columnwidth]{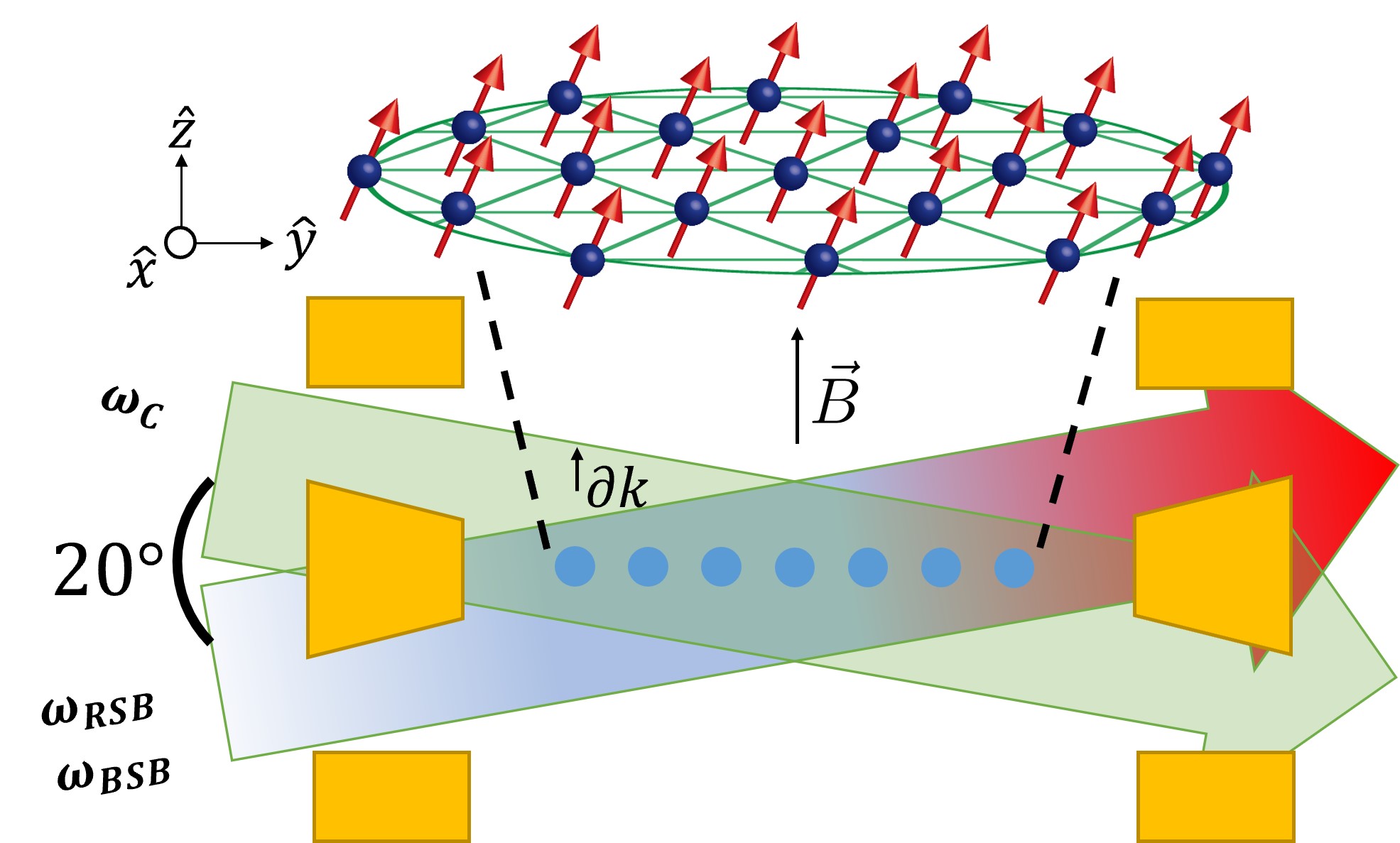}
    \caption{Schematic of the lasers required for driving a M\o lmer-S\o rensen gate in the Penning trap. The beam with angular frequency $\omega_C$ is assumed to be the upper beam, while the red and blue sidebands (angular frequencies $\omega_{RSB}$ and $\omega_{BSB}$, respectively) copropagate along the lower beam path. The relationship between $\omega_C$ and $\omega_{RSB(BSB)}$ depends on whether the ``phase-sensitive" or ``phase-insensitive" configuration is used. The angle between the beams is assumed to remain the same as in the LS gate (see Fig.~\hyperref[fig:BeLevelsODFBeams]{\ref*{fig:BeLevelsODFBeams}(b)}).}
    \label{fig:MS_Lasers_Penning}
\end{figure}

While previous NIST Penning trap experiments used the LS gate discussed in Sec.~\ref{sec:LSGateOptimization} \cite{Britton2012,bohnet2016,gilmore2021,garttner2017}, there may be advantages to using the MS gate. Depending on the experimental configuration, the MS gate can increase the robustness to relative phase fluctuations of the laser beams involved compared with the LS gate. The MS gate when driven off-resonantly on an electric dipole transition requires a minimum of three laser beams\,---\,a beam, frequently denoted as the carrier \cite{haljan2005}, which has angular frequency $\omega_C$, and the red and blue sidebands (RSB and BSB respectively), which have angular frequencies of $\omega_{RSB}$ and $\omega_{BSB}$. There are multiple options for the frequencies of the red and blue sideband laser beams and for the spatial configuration of the beams. We will assume here that both the red and blue sidebands are applied with the lower laser beam and the carrier with the upper laser beam as shown in Fig.~\ref{fig:MS_Lasers_Penning}. One possible frequency configuration is $\omega_{BSB}= \omega_{C}+\Delta_{\uparrow\downarrow} +\omega_z + \delta_z$ and $\omega_{RSB} = \omega_{C} + \Delta_{\uparrow\downarrow}-\omega_z - \delta_z$. Another possibility entails detuning the red and blue sidebands in opposite directions from the carrier frequency. Specifically, in this case, $\omega_{BSB} = \omega_{C} + \Delta_{\uparrow\downarrow} + \omega_z + \delta_z$ as before but $\omega_{RSB} = \omega_{C} - \Delta_{\uparrow\downarrow} + \omega_z + \delta_z$ \cite{haljan2005}. The first beam geometry is known as the ``phase-sensitive" configuration, while the second is known as the ``phase-insensitive" configuration. These names refer only to the phase on the spin of the ions, however, and not the phase on the motional state during the gate. The phase-insensitive gate thus results in a qubit spin state (see Eq. 1 of Ref.~\cite{haljan2005}) that is robust to differential phase fluctuations in the red and blue sidebands, but a motional state that is sensitive to these differential phase fluctuations. The phase-sensitive gate has the opposite effect. For both the sensing work discussed in Refs.~\onlinecite{gilmore2021} and \onlinecite{affolter2020} and driving gates in 3D crystals, we require the motional state to be insensitive to phase fluctuations, and are thus interested in the so-called phase-sensitive scheme. In this beam geometry, we would obtain the phase stability and other advantages discussed here and in Sec.~\ref{sec:Introduction}.

\begin{figure*}
    \centering
    \includegraphics[width=\textwidth]{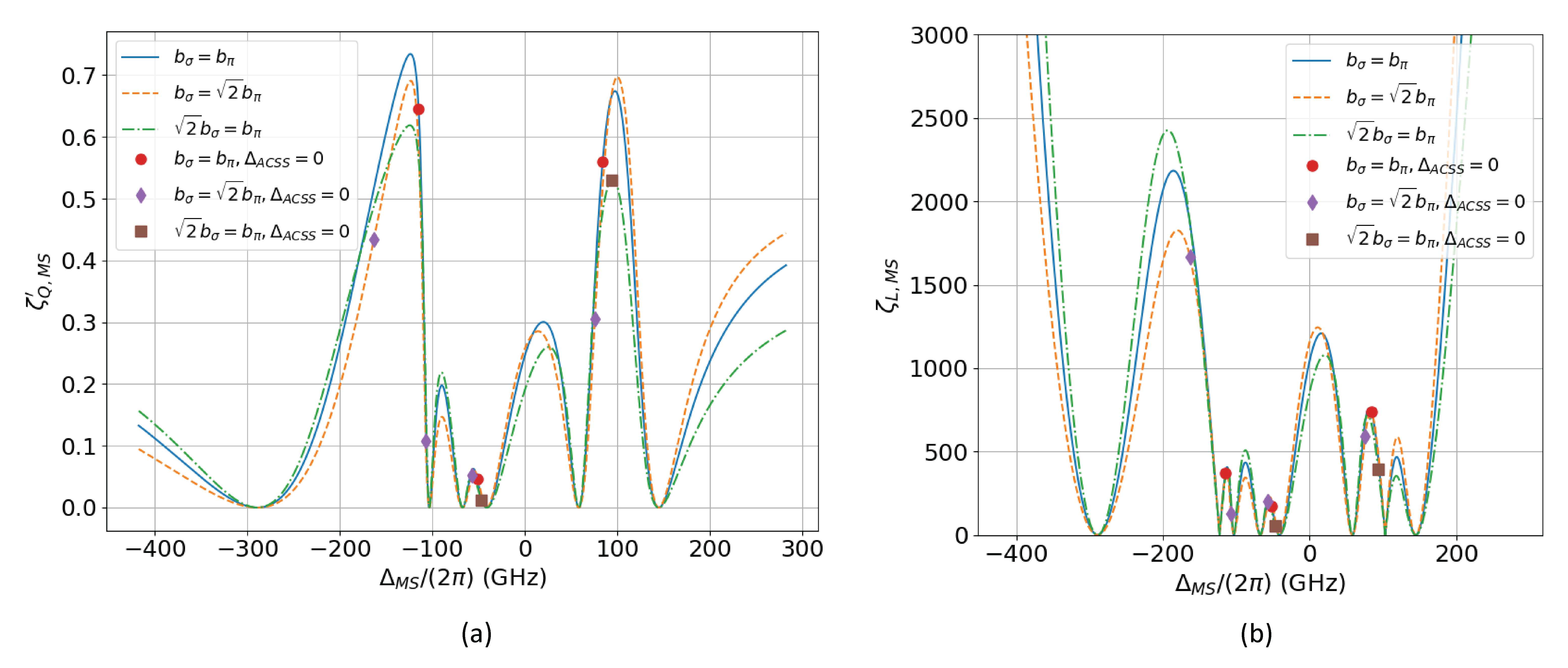}
    \caption{Figures of merit for the MS gate versus $\Delta_{MS}/(2\pi)$ for configuration 1. We show three possible settings of the relative electric field amplitudes of the two laser beams\,---\,$b_\sigma = b_\pi$ (blue solid line), $b_\sigma = \sqrt{2}b_\pi$ (orange dashed line) and $\sqrt{2}b_\sigma = b_\pi$ (green dashed and dotted line). Additionally, we indicate the detunings where these intensity settings result in $\Delta_{ACSS}=0$ with red circles, purple diamonds, and brown squares, respectively. (a) $\zeta^\prime_{Q, MS}$ versus detuning $\Delta_{MS}$. (b) $\zeta_{L, MS}$ versus detuning $\Delta_{MS}$. The far-detuning regime where $\zeta_{L, MS}$ diverges has similarly stringent power requirements to the comparable regime for the LS gate. However, the strong suppression of spontaneous emission makes consideration of this regime important. Further details on operating in this regime are discussed in Sec.~\ref{sec:OperatingPoints}.}
    \label{fig:MSFOM}
\end{figure*}

However, if the MS gate were to introduce significantly more decoherence due to spontaneous emission than the LS gate, that could limit its utility for quantum simulations and sensing because decoherence due to off-resonant light scatter already is the leading source of error in the current operating configuration at NIST (see Sec.~\ref{sec:LSBackground} and Ref.~\cite{bohnet2016}). Thus, we want to be able to compare $\zeta_{Q, LS}$ and $\zeta_{L, LS}$ with equivalent figures of merit $\zeta_{Q, MS}$ and $\zeta_{L, MS}$ for the MS gate. Additionally, we would ideally like to maintain the condition that the differential ACSS is nulled. We consider two possibilities\,---\, one where we require that the differential AC Stark shift be nulled ($\Delta_{ACSS} = 0$) when all beams are applied (configuration 1) and a second where the AC Stark shift is nulled for each individual laser beam (configuration 2).

For the following calculations, we will define $\Delta_{MS} \equiv \frac{1}{2}\left(\omega_{RSB} + \omega_{BSB}\right) - \omega_0 = \omega_C + \Delta_{\uparrow\downarrow} - \omega_0$ where $\hbar\omega_0$ was previously defined as the zero-field frequency difference between the $S_{1/2}$ and $P_{3/2}$ states. Additionally, for the characterization of the ACSS and spontaneous
emission, we will neglect the frequency differences in the
red and blue sidebands since $\omega_z, \delta_z \ll \Delta_{\uparrow\downarrow}$. Specifically, we will take $\omega_{SB} = \omega_{RSB} = \omega_{BSB} = \omega_C + \Delta_{\uparrow\downarrow}$ when calculating the detunings from atomic resonances for the scattering amplitudes and the ACSS. This assumption does not affect the calculation of the single-spin decoherence rate, as discussed further in Appendix~\ref{sec:AppA}.

As indicated above, we consider two laser configurations, which are summarized in Table~\ref{tab:MS_config}. First, we will assume the sideband (higher frequency) beam has pure horizontal polarization, while the carrier (lower frequency) beam has pure vertical polarization. With the beams aligned approximately perpendicularly to the magnetic field, this results in the sidebands having equal components of $\sigma^+$ and $\sigma^-$ polarized light but no $\pi$ polarized light, while the carrier has only $\pi$ polarization. In the second configuration, we will consider the case where the two beams have the same intensity, but we select the polarization angles of each beam such that either beam results in a nulled differential ACSS. 

For configuration 1 (see Table~\ref{tab:MS_config}), we will allow the relative intensities of the two beams to vary while maintaining a constant total laser power. We define the electric fields of the two beams,
\begin{align}
    \vec{\mathcal{E}}_{C_1} = \mathcal{E}_{C_0}\hat{z}\cos\left(\vec{k}_C\cdot\vec{r}-\omega_{C}t\right),\nonumber\\
    \vec{\mathcal{E}}_{SB_1} = \mathcal{E}_{SB_0}\hat{x}\cos\left(\vec{k}_{SB}\cdot\vec{r}-\omega_{SB}t\right), \label{eq:Efield_MS_perp}
\end{align}
where we are making the same small angle approximation as for the polarization of the beams in the LS gate. To ensure correct normalization of the intensity, we impose the constraint $\mathcal{E}_{C_0}^2 + \mathcal{E}_{SB_0}^2=2\mathcal{E}_{MS_0}^2$. We can thus parameterize the relative intensities with a single effective angle $\varphi_{MS}$ with $\mathcal{E}_{SB_0}=\sqrt{2}\cos\left(\varphi_{MS}\right)\mathcal{E}_{MS_0}$ and $\mathcal{E}_{C_0} = \sqrt{2}\sin\left(\varphi_{MS}\right)\mathcal{E}_{MS_0}$. 

For configuration 2, as described in Table~\ref{tab:MS_config}, the two beams have equal intensities, and the polarizations are set such that $\Delta_{ACSS}=0$ with either beam. The electric fields are given by
\begin{align}
    \vec{\mathcal{E}}_{C_2} &= \mathcal{E}_{MS_0}\hat{\epsilon}_{C}\cos\left(\vec{k}_C\cdot\vec{r} - \omega_C t\right),\nonumber\\
    \vec{\mathcal{E}}_{SB_2} &= \mathcal{E}_{MS_0}\hat{\epsilon}_{SB}\cos\left(\vec{k}_{SB}\cdot\vec{r}-\omega_{SB}t\right), \label{eq:Efield_MS_null}
\end{align}
where, assuming linear polarization,
\begin{align}
    \hat{\epsilon}_C &= \cos\left(\phi_{C}\right)\hat{z} + \sin\left(\phi_C\right)\hat{x},\nonumber\\
    \hat{\epsilon}_{SB} &= \cos\left(\phi_{SB}\right)\hat{z} + \sin\left(\phi_{SB}\right)\hat{x}.
\end{align}
\begin{table}[t]
    \renewcommand{\arraystretch}{1.5}
    \centering
    \begin{tabular}{|>{\centering\arraybackslash}m{.11\columnwidth}|>{\raggedright\arraybackslash}p{.38\columnwidth}|>{\centering\arraybackslash}m{.2\columnwidth}|>{\centering\arraybackslash}m{.17\columnwidth}|}
    \hline
    \textbf{Label} & \multicolumn{1}{c|}{\textbf{Description}} & \textbf{Electric field equations} & \textbf{Figure of merit plots}\\
    \hline
    1 & Perpendicularly polarized beams, $\sigma$ polarization in high-frequency beam and $\pi$ polarization in low-frequency beam. Relative intensities can vary. & Eq.~\ref{eq:Efield_MS_perp} & \makecell{\\Fig.~\ref{fig:MSFOM}\\ Fig. \ref{fig:MSmultiB}}\\
    \hline
    2 & Polarizations set such that $\Delta_{ACSS}=0$ for each beam. Intensities are the same for both beams. & Eq.~\ref{eq:Efield_MS_null} & Fig.~\ref{fig:MSFOMnull}\\
    \hline
    \end{tabular}
    \caption{Summary of the configurations considered for the M\o lmer-S\o rensen gate with a brief description and references to the electric field definitions and plots of the figures of merit.}
    \label{tab:MS_config}
\end{table}
In principle, these electric fields could have any linear polarization, but we will limit our discussion here to the case where $\phi_C$ and $\phi_{SB}$ are set to null the ACSS from $\vec{\mathcal{E}}_{C_2}$ and $\vec{\mathcal{E}}_{SB_2}$ respectively. 

\subsection{Spontaneous emission in the M\o lmer-S\o resen gate}\label{sec:MSSpontaneousEmission}

Given these laser parameters, we now look at the calculation of single-qubit decoherence due to spontaneous emission in the MS gate. This calculation is less straightforward than that for the LS gate because of the different frequencies of the carrier and sideband laser beams. Also, the inclusion of the appropriate coherent interaction, which for the MS gate is a $\hat{\sigma}^x$ interaction, changes the expression for the single-qubit decoherence from that used for the LS gate (Eq.~\ref{eq:Gamma_LS}). Further details are discussed in Appendix~\ref{sec:AppA}. We find that the single-qubit decoherence rate due to spontaneous emission in the MS gate is
\begin{equation}
    \Gamma_{MS} = \frac{1}{4}\left(\Gamma_{el} + 3\Gamma_r\right),
    \label{eq:Gamma_MS}
\end{equation}
where $\Gamma_{el}$ and $\Gamma_r$ are the rates of decoherence due to Rayleigh and Raman scattering from all laser beams (the carrier as well as the sideband beams) obtained from the methodology of Ref.~\onlinecite{Uys2010} and calculated using Eqs.~\ref{eq:Gamma_ij_singlebeam}, \ref{eq:Gamma_r_ZZ_singlebeam}, and \ref{eq:Gamma_el_ZZ_singlebeam} for each beam and adding the single beam rates together. This rate can be compared to $\Gamma$ for the LS gate (Eq.~\ref{eq:Gamma_LS}).

\subsection{Interaction strength and figure of merit for the M\o lmer-S\o rensen gate}\label{sec:MSFOM}

To determine the figures of merit for the MS gate, $\zeta_{Q,MS}$ and $\zeta_{L, MS}$, we first consider the Ising interaction strength $J_{ij}$, or if we couple primarily to the COM mode, $J$. For the MS gate, $J$ can be written in terms of experimental parameters as \cite{kim2009}
\begin{equation}
    J\approx \frac{\hbar \left(\Omega_R \partial k\right)^2}{4 m\omega_z \delta_z},
    \label{eq:J_XX}
\end{equation}
where $\Omega_R$ is the two-photon Rabi frequency. Since we are assuming the carrier beam has a lower frequency than the sideband beam, $\Omega_R$ can be written as
\begin{widetext}
\begin{align}
    \Omega_R &= \left(\frac{g_0}{\mu}\right)^2\left(\frac{\Braket{\uparrow| \vec{d}\cdot \vec{\mathcal{E}}'_{C_i}|3}\Braket{3|\vec{d}\cdot\vec{\mathcal{E}}'_{SB_i}|\downarrow}}{\Delta_{MS} +\left(\omega_0 - \omega_3\right)} + \frac{\Braket{\uparrow|\vec{d}\cdot\vec{\mathcal{E}}'_{C_i}|6}\Braket{6|\vec{d}\cdot\vec{\mathcal{E}}'_{SB_i}|\downarrow}}{\Delta_{MS} + \left(\omega_0 -\omega_6\right)}\right. \nonumber\\
    &\left. + \frac{\Braket{\uparrow| \vec{d}\cdot \vec{\mathcal{E}}'_{C_i}|2}\Braket{2|\vec{d}\cdot\vec{\mathcal{E}}'_{SB_i}|\downarrow}}{\Delta_{MS} +\left(\omega_0 - \omega_2\right)} + \frac{\Braket{\uparrow |\vec{d}\cdot\vec{\mathcal{E}}'_{C_i}|5}\Braket{5|\vec{d}\cdot\vec{\mathcal{E}}'_{SB_i}|\downarrow}}{\Delta_{MS} +\left(\omega_0 - \omega_5\right)}  \right),
    \label{eq:Omega_R}
\end{align}
\end{widetext}
where $\omega_j$ is defined as in Eq.~\ref{eq:ACSS_coeffs} and $g_0=\left(\mu\mathcal{E}_{MS_0}\right)/(2\hbar)$. Additionally, $i$ gives the index for which configuration of laser beams (either configuration 1 or configuration 2 from Table~\ref{tab:MS_config}) is used and $\vec{\mathcal{E}}'_{C_i}$ and $\vec{\mathcal{E}}'_{SB_i}$ equal $\vec{\mathcal{E}}_{C_i}/\mathcal{E}_{MS_0}$ and $\vec{\mathcal{E}}_{SB_i}/\mathcal{E}_{MS_0}$, respectively.

\begin{figure*}[t]
    \centering
    \includegraphics[width=\textwidth]{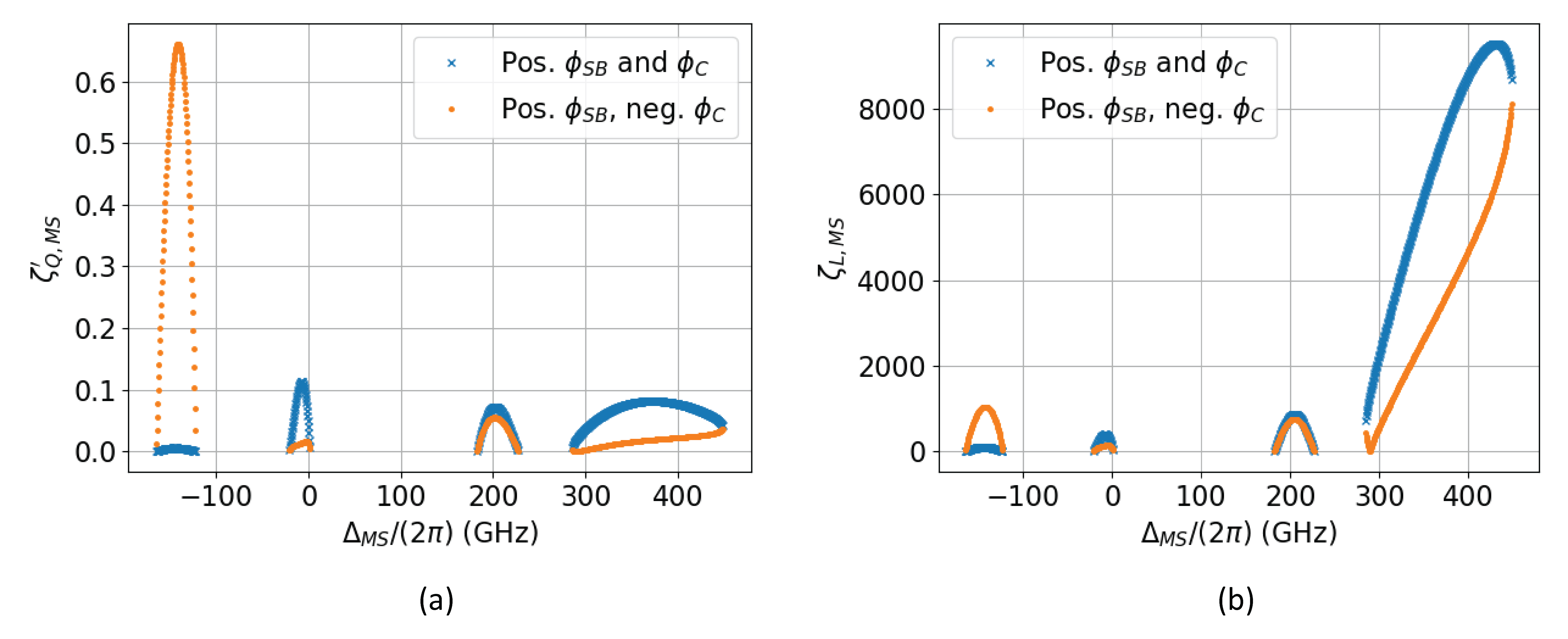} 
    \caption{Figures of merit for the MS gate versus detuning for configuration 2 (Table~\ref{tab:MS_config}). The points where both $\phi_{SB}$ and $\phi_C$ are positive are shown with blue x's and the points where $\phi_{SB}$ is positive and $\phi_C$ is negative are shown with orange circles.}
    \label{fig:MSFOMnull}
\end{figure*}

By comparing Eq.~\ref{eq:J_XX} with Eq.~\ref{eq:JCOM} and considering the LS gate figures of merit as defined in Eqs.~\ref{eq:FOM_ZZ} and \ref{eq:LS_LinFOM}, we see that appropriate corresponding figures of merit for the MS gate are
\begin{align}
    \zeta_{Q, MS} &= \frac{\Omega_R^2}{\Gamma_{MS}},\label{eq:FOM_XX}\\
    \zeta_{L, MS} &= \frac{\Omega_R}{\Gamma_{MS}}.\label{eq:MS_LinFOM}
\end{align}

As with the LS gate, we define a scaled quadratic figure of merit $\zeta^\prime_{Q, MS}=\zeta_{Q, MS}/\zeta_{Q,0}$. We plot both figures of merit for configuration 1 (see Table~\ref{tab:MS_config}) in Fig.~\ref{fig:MSFOM}. We consider various values of $\varphi_{MS}$, or equivalently $b_\sigma$ and $b_\pi$, where $b_\sigma = \sqrt{2}\sin\left(\varphi_{MS}\right)$ and $b_\pi =\sqrt{2}\cos\left(\varphi_{MS}\right)$. We indicate points where the differential ACSS is nulled for increased robustness to laser intensity fluctuations. In Sec.~\ref{sec:OperatingPoints}, we will compare various operating points for the linear figure of merit for both configuration 1 and configuration 2 for the MS gate with various operating points for the LS gate. For now, we note that the maximum values for $\zeta^\prime_{Q,MS}$ are generally comparable to those for $\zeta^\prime_{Q,LS}$ when the ACSS is nulled although slightly smaller. For the linear figure of merit, the point at a detuning of approximately -160~GHz in configuration 1 with $b_\sigma = \sqrt{2}b_\pi$ (shown in purple in Fig.~\hyperref[fig:MSFOM]{\ref*{fig:MSFOM}(b)}) yields a value of $\zeta_{L,MS}$ that is about 1.5 times larger than the value of $\zeta_{L,LS}$ at the current NIST operating point for the LS gate.

Additionally, we can consider the resulting figures of merit when the applied electric fields are given by $\vec{\mathcal{E}}_{C_2}$ and $\vec{\mathcal{E}}_{SB_2}$ as defined in Eq.~\ref{eq:Efield_MS_null}. The advantage of this configuration (configuration 2 in Table~\ref{tab:MS_config}) is that it permits nulling of the differential ACSS at many different detunings and reduces sensitivity to intensity fluctuations that differ in the two beams. In principle, we could also consider allowing the relative amplitudes to vary as we did for configuration 1 (Eq.~\ref{eq:Efield_MS_perp}), but for the sake of simplicity and providing an example, we consider only the case where the amplitudes are both equal.

We plot the resulting figures of merit in Fig.~\ref{fig:MSFOMnull}. For each detuning, if there is a polarization angle $\phi$ that nulls the differential ACSS, then both $+\phi$ or $-\phi$ will null the differential ACSS. Changing the sign of the polarization angle for one beam results in changing the sign of two of the terms in $\Omega_R$ (Eq.~\ref{eq:Omega_R}), which can change constructive interference to destructive and vice versa. We note that the maximum value of $\zeta^\prime_{Q, MS}$ in configuration 2 is comparable to that in configuration 1. Also, it is worth noting that there is an operating point at $\Delta_{MS}/(2\pi)=-142.6$~GHz in configuration 2 with a value of $\zeta_{L,MS}$ comparable to that for $\zeta_{L,LS}$ at the current NIST operating point (compare Fig.~\hyperref[fig:MSFOMnull]{\ref*{fig:MSFOMnull}(b)} with Fig.~\hyperref[fig:FOMZZDeltaACSS=0]{\ref*{fig:FOMZZDeltaACSS=0}(b)}). More details of potential operating points for the MS gate will be discussed in Sec.~\ref{sec:OperatingPoints}.

\begin{figure*}[t]
    \centering
    \includegraphics[width=\textwidth]{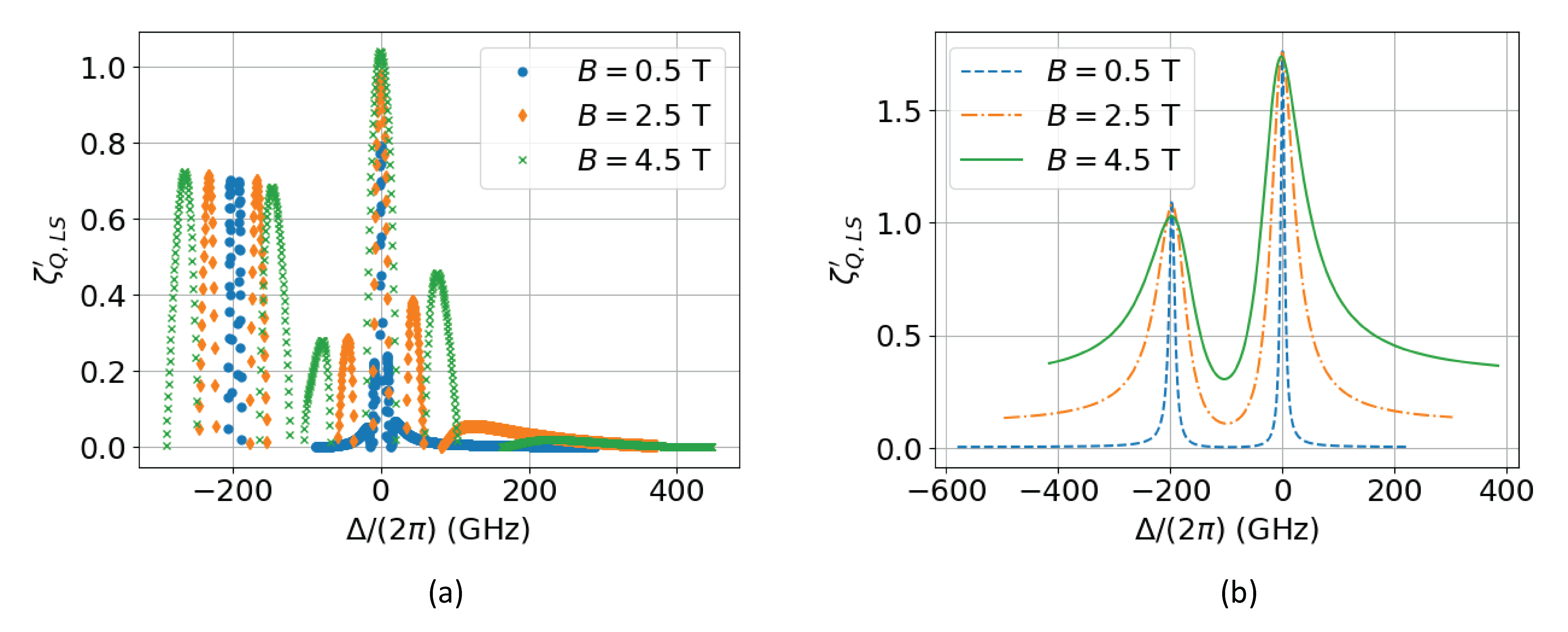}
    \caption{Plots of the quadratic figure of merit for magnetic fields of 0.5~T, 2.5~T, and 4.5~T versus $\Delta$. (a) $\zeta^\prime_{Q, LS}$ assuming the ACSS is nulled. The 0.5~T points are shown with blue circles, the 2.5~T points are shown with orange diamonds, and the 4.5~T points are shown with green x's. The maximum point is at approximately the same frequency for every field, but the figure of merit slowly decreases as the field decreases. (b) $\zeta^\prime_{Q, LS}$ for vertical polarization at 0.5~T (blue dashed line), 2.5~T (orange dashed and dotted line), and 4.5~T (green solid line). Not only does the optimal operating frequency remain approximately constant, but the maximum figure of merit remains approximately constant as well. As discussed when introducing the quadratic figure of merit (see Sec.~\ref{sec:FOM}), all curves assume the same fixed laser intensity.}
    \label{fig:FOMZZMultiB}
\end{figure*}

\section{Gate performance for varying magnetic fields}\label{sec:VaryingB}
While we currently operate our experiment at $\sim$4.5~T, this field is not necessarily the most favorable. In particular, using Penning traps at lower magnetic fields has some technical advantages, including the possibility of employing permanent magnets resulting in a much more compact setup \cite{ball2019, mcmahon2020}. We explore the viability of both the LS and MS gates at lower fields.

In Fig.~\hyperref[fig:FOMZZMultiB]{\ref*{fig:FOMZZMultiB}(a)} we plot $\zeta^\prime_{Q, LS}$ versus $\Delta$ for points where $\Delta_{ACSS}=0$ for magnetic fields of 0.5~T, 2.5~T and 4.5~T. There is a small decrease in the maximum figure of merit from 4.5~T to 2.5~T and then a larger decrease at 0.5~T. However, if we relax the constraint on the differential ACSS and consider vertically polarized beams, the maximum value of the figure of merit remains approximately constant for all fields as shown in Fig.~\hyperref[fig:FOMZZMultiB]{\ref*{fig:FOMZZMultiB}(b)}. Additionally, it is interesting to note that the optimal laser frequency for both configurations remains constant regardless of magnetic field, corresponding to the zero-field frequency difference between the $S_{1/2}$ and $P_{3/2}$ manifolds.. The second peak at a detuning $\Delta/(2\pi) \approx -200$~GHz in Fig.~\hyperref[fig:FOMZZMultiB]{\ref*{fig:FOMZZMultiB}(b)} corresponds to a laser frequency approximately equal to the zero-field frequency difference between the $S_{1/2}$ and $P_{1/2}$ levels. 

\begin{figure}
    \centering
    \includegraphics[width=.85\columnwidth]{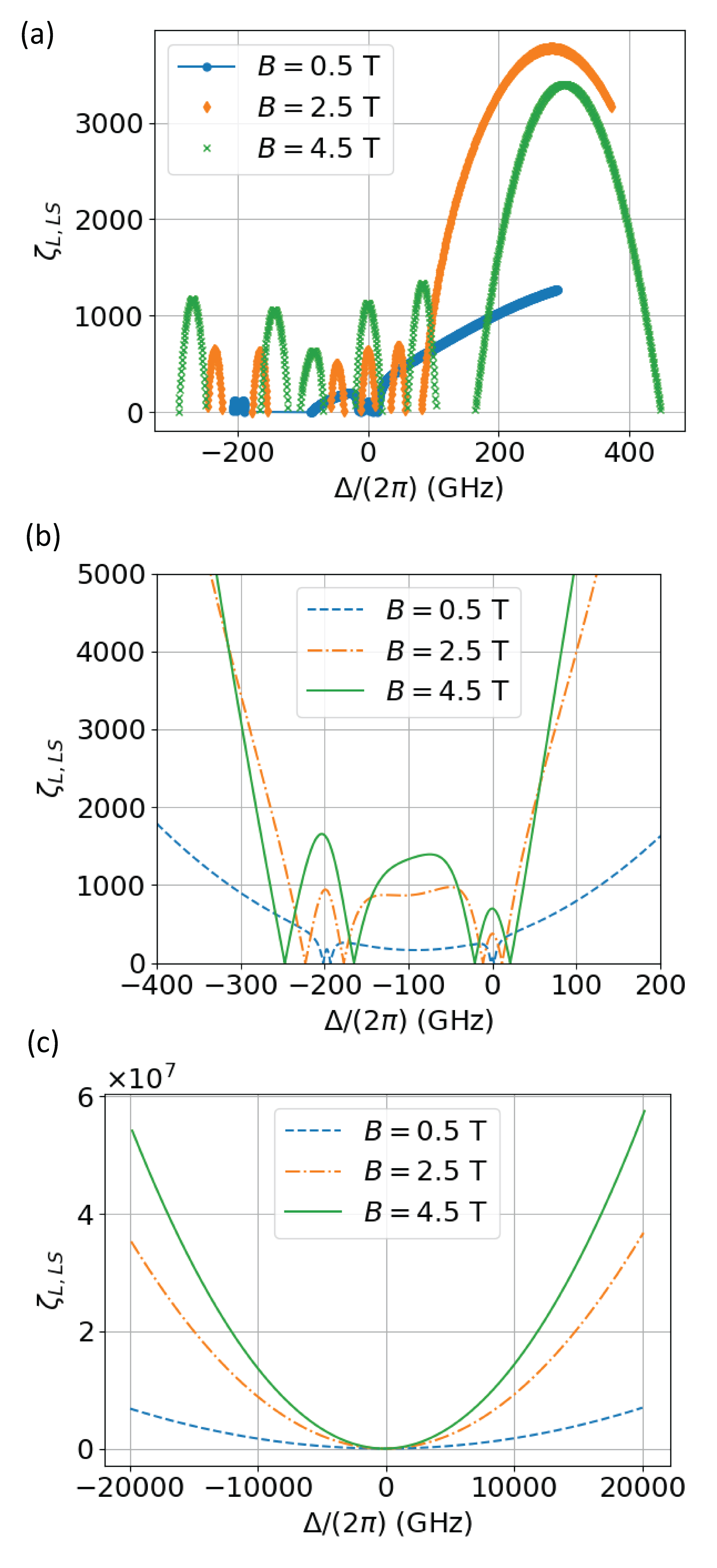}
    \caption{Plots of $\zeta_{L, LS}$ for magnetic fields of 0.5~T, 2.5~T, and 4.5~T versus $\Delta$. (a) $\zeta_{L, LS}$ assuming the ACSS is nulled for 0.5~T (blue circles), 2.5~T (orange diamonds), and 4.5~T(green x's). (b) $\zeta_{L, LS}$ with vertical polarization for 0.5~T (blue dashed line), 2.5~T (orange dashed and dotted line), and 4.5~T (green solid line). (c) $\zeta_{L,LS}$ for 0.5~T (blue dashed line), 2.5~T (orange dashed and dotted line), and 4.5~T (green solid line) assuming vertically polarized laser beams for large values of $\Delta$.}
    \label{fig:FOMZZlinmultiB}
\end{figure}

\begin{figure*}[t]
    \centering\includegraphics[width=\textwidth]{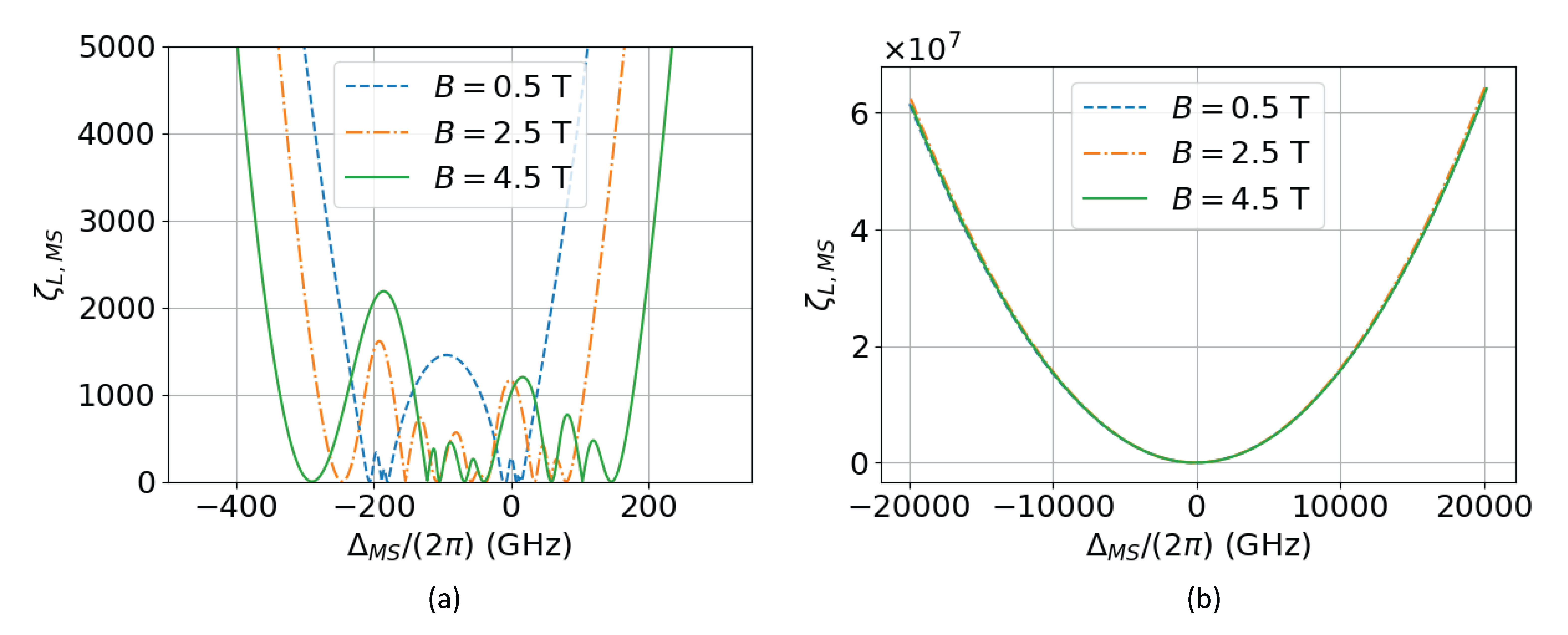}
    \caption{Plots of linear figures of merit for the MS gate for magnetic fields of 0.5~T (blue dashed line), 2.5~T (orange dashed and dotted line), and 4.5~T (green solid line). We assume the beams are configured according to configuration 1 (see Table~\ref{tab:MS_config}) and have equal intensities. (a) $\zeta_{L,MS}$ vs $\Delta_{MS}$ for detunings near atomic transitions. (b) $\zeta_{L,MS}$ vs $\Delta_{MS}$ for large $\Delta_{MS}$. The value of $\zeta_{L,MS}$ is nearly constant with magnetic field so the curves for the three fields are nearly on top of each other.}
    \label{fig:MSmultiB}
\end{figure*}

In the quantum simulation regime, tuning between the Zeeman levels appears to provide a viable operating point even for low magnetic fields. For the gate regime, Figs.~\hyperref[fig:FOMZZlinmultiB]{\ref*{fig:FOMZZlinmultiB}(a)} and \hyperref[fig:FOMZZlinmultiB]{(b)} show plots of the linear figure of merit for magnetic fields of 0.5~T, 2.5~T, and 4.5~T both for when the ACSS can be nulled and for vertical polarization. The linear figure of merit $\zeta_{L,LS}$ is lower for weaker magnetic fields when the laser frequencies are tuned between the excited state Zeeman levels. Qualitatively, from Eqs.~\ref{eq:ACSS_coeffs}, \ref{eq:Gamma_ij_singlebeam}, and \ref{eq:ScatteringAmplitudes} one expects $\zeta_{L,LS}$ to increase with $\Delta$. Because the spacing between the Zeeman levels decreases with decreasing magnetic field, tuning between the Zeeman levels results in decreasing performance as the magnetic field is reduced. 

In Fig.~\hyperref[fig:FOMZZlinmultiB]{\ref*{fig:FOMZZlinmultiB}(c)} we further demonstrate this decrease in performance of the LS gate linear figure of merit with decreasing magnetic field, but now in a regime where $\Delta$ is large compared to all other relevant frequencies. This reinforces the trend preliminarily observed in Fig.~\hyperref[fig:FOMZZlinmultiB]{\ref*{fig:FOMZZlinmultiB}(b)}, where $\zeta_{L,LS}$ increases more rapidly with detuning as the magnetic field increases. We can understand the magnetic field dependence of $\zeta_{L,LS}$ by considering the spin-dependent force and rate of decoherence due to spontaneous emission independently. The spin-dependent force is proportional to the differential AC Stark shift between the levels $\ket{\downarrow}$ and $\ket{\uparrow}$. For the LS gate with linearly polarized laser beams (see Sec.~\ref{sec:LSBackground}), the matrix elements that couple the $\ket{\downarrow}$ and $\ket{\uparrow}$ levels to the excited $P$ states are identical, and a differential AC Stark shift is obtained only through different detunings in the denominators in the expressions for the ACSS for $\ket{\downarrow}$ and $\ket{\uparrow}$ in Eq.~\ref{eq:ACSS_coeffs}. At zero magnetic field, the detunings are the same for both qubit states, resulting in no spin-dependent force. As the magnetic field $B$ increases, the difference in the detunings for $\ket{\downarrow}$ and $\ket{\uparrow}$ increases, resulting in a spin-dependent force that increases linearly with magnetic field. For large detunings $\Delta$, one can show that the spin-dependent force increases linearly with the ratio of the Zeeman splittings (parameterized as $\Delta_z = \frac{\mu_B}{\hbar}B$) to the square of the detuning $\Delta$ (see Appendix~\ref{appendix:BomegaFS_dependence} and Eq.~\ref{eq:F0approx}). $\Gamma_r$ and $\Gamma_{el}$ are both independent of the magnetic field strength to lowest order in $\Delta_z/\Delta$.

LS gates could be improved by having the ODF laser beams enter the trap at larger angles with respect to the plane of the ion crystal, thus enabling significantly different intensities for $\sigma^+$ and $\sigma^-$ polarized light. This laser beam configuration would result in a nonzero differential ACSS even at zero magnetic field because of the asymmetry between transitions driven by the different $\sigma$ polarizations. This would mitigate the strong magnetic field dependence of $\zeta_{L,LS}$ obtained for linear polarizations with beams perpendicular to $\vec{B}$. As briefly discussed, larger angle ODF beams will generate a shorter wavelength 1D optical lattice, which can increase the challenge of aligning the ODF wavefronts with the extended single-plane crystal.

We can also look at whether the MS gate has a similar magnetic field dependence. We plot $\zeta_{L,MS}$ vs $\Delta_{MS}/(2\pi)$ for a variety of fields in Fig.~\ref{fig:MSmultiB} for configuration 1 (see Table~\ref{tab:MS_config}). The plot in Fig.~\hyperref[fig:MSmultiB]{\ref*{fig:MSmultiB}(a)} shows that, while the optimal values of $\Delta_{MS}$ shift depending on the magnetic field, there are operating points for every magnetic field shown that have values of $\zeta_{L,MS}$ comparable to the value of $\zeta_{L,LS}$ for the current NIST operating point. Additionally, we demonstrate with the plot in Fig.~\hyperref[fig:MSmultiB]{\ref*{fig:MSmultiB}(b)} that the linear figure of merit does not, in fact, decrease significantly with decreasing magnetic field for the MS gate for large detunings. This result is in agreement with the discussion in Appendix~\ref{appendix:BomegaFS_dependence}, which shows that there is no leading order dependence of $\Omega_R$ or $\Gamma_{MS}$ on the magnetic field (i.e. on $\Delta_z/\Delta_{MS}$). This analysis suggests that the MS gate may be preferable to the LS gate at lower fields when laser beams nearly perpendicular to the magnetic field are desired.

We note that the scaling of the linear figures of merit for the LS and MS gates with the detuning $\Delta$ and $\Delta_{MS}$, respectively appears to be quadratic for large detunings. While it may be more intuitive to assume that $F_0 \propto 1/\Delta$, $\Omega_R \propto 1/\Delta_{MS}$, $\Gamma \propto 1/\Delta^2$, and $\Gamma_{MS} \propto 1/\Delta_{MS}^2$ (see Eqs.~\ref{eq:ACSS_coeffs}, \ref{eq:Gamma_ij_singlebeam}, \ref{eq:ScatteringAmplitudes}, \ref{eq:Omega_R}, and \ref{eq:Gamma_MS}), the more detailed analysis of Appendix~\ref{appendix:BomegaFS_dependence} shows that, due to the interaction with both the $P_{1/2}$ and $P_{3/2}$ levels, the interaction strengths $F_0$ and $\Omega_R$ are proportional to $1/\Delta^2$ and $1/\Delta_{MS}^2$ respectively. Additionally, $\Gamma\propto 1/\Delta^4$ and $\Gamma_{MS} \propto 1/\Delta_{MS}^4$, resulting in an overall scaling for the linear figures of merit for the LS and MS gates with $\Delta^2$ and $\Delta_{MS}^2$ respectively, assuming large detunings.

\section{Spontaneous emission errors in a two-qubit light-shift gate}\label{sec:Fidelity} 
The analysis of Secs.~\ref{sec:LSGateOptimization} and \ref{sec:MSGateHighFields} derives optimal laser detunings $\Delta \left(\Delta_{MS}\right)$ for implementing the LS (MS) gate in a high magnetic field. Tuning the optical dipole force laser beams to a virtual level between the Zeeman levels, as is implemented in the NIST Penning trap, is shown to have local maxima in the linear figures of merit. Additionally, setting the detuning $\Delta$ between the Zeeman levels results in useful features such as AC Stark shift nulls and modest laser power requirements. These detuning settings become possible only in strong magnetic fields. A careful comparison between the potential performance that can be achieved at high magnetic field by tuning between the Zeeman levels and at low magnetic field, where trapped ion gates are performed with large detunings from any excited state levels, is of interest. A frequent assessment of the quality of a quantum information system is the fidelity of a maximally entangling gate between two qubits. While in practice we operate with many more than two qubits, this example will serve as a useful point of comparison with other quantum systems. 

A previous characterization of the fidelity limitations of a two-qubit gate with spontaneous emission in trapped ion clock qubits was discussed in Ref.~\onlinecite{ozeri2007}. However, that paper does not consider the direct impacts of Rayleigh scattering on the ion coherence, instead considering only the impact of recoil from Rayleigh scattering. This approach is valid for clock qubits, where $\sum_{J}A_{J,\lambda}^{\uparrow\rightarrow\uparrow}$ and $\sum_{J'}A_{J',\lambda}^{\downarrow\rightarrow\downarrow}$ are nearly equal for all $\lambda$ for typical experimental parameters. However, for the Zeeman qubits discussed here, elastic scattering contributes directly to the decoherence of the internal state of the ion. 

Although general methods for calculating the rates of decoherence from off-resonant light scatter have been established in the literature \cite{Uys2010}, an expression for the fidelity of a two-qubit entangling gate in the presence of off-resonant light scatter is not available. In this section, we present such an expression for the LS gate and evaluate it for relevant operating configurations. Appendix~\ref{appendix:fidelity} details the derivation of the expression and a comparable result for the MS gate.

To derive this fidelity expression, we assume the qubits are prepared in the state $\ket{\psi_i} = \ket{++}$ where $\ket{\pm} = \frac{1}{\sqrt{2}}\left(\ket{\downarrow} \pm \ket{\uparrow}\right)$. We then allow the qubits to evolve under $\hat{H}_{LS}$ from Eq.~\ref{eq:Ising_general} with $\mathcal{N}$=2 and $J_{ij}\rightarrow J$ since there is a single pair of ions involved. The resulting ideal state is
\begin{equation}
    \Ket{\psi_0}=  \frac{1}{2}\sum_{\sigma_1^z \sigma_2^z = \pm 1}e^{-\frac{1}{2}i J t\sigma_1^z \sigma_2^z}\Ket{\sigma_1^z \sigma_2^z},
    \label{eq:IdealBellZZ}
\end{equation}
where we parameterize the qubit states of the $i^{th}$ ion with the numbers $\sigma_i^z = \pm 1$ corresponding to the qubit states $\ket{\uparrow}$ and $\ket{\downarrow}$ respectively. When $t=\tau_g = \frac{\pi}{2J}$, this state corresponds to a maximally entangled Bell state.

To find the fidelity, we compute the overlap between the density matrix for the ideal Bell state \begin{equation}
    \Ket{\Psi_b} = \frac{1}{\sqrt{2}}\left(\ket{++} - i\ket{--}\right),
\end{equation}
with the density matrix of the two-ion system at time $\tau_g = \pi/(2J)$ accounting for spontaneous emission. The derivation of the density matrix with spontaneous emission and the resulting full expression are presented in Appendix~\ref{appendix:fidelity}. Assuming that the interaction strength $J$ is large compared to $\Gamma_r$ and $\Gamma_{el}$, we obtain a simplified expression for the fidelity at the Bell state time
\begin{equation}
    \mathcal{F}_{LS}\left(\tau_g\right)\approx 1-\left(\frac{3}{4}\Gamma_r + \frac{1}{2}\Gamma_{el}\right)\tau_g
    \label{eq:Fidelity_ZZ_approx}
\end{equation}
as shown in Appendix~\ref{appendix:fidelity}.

A key feature of this result is the suppression of the impact of Rayleigh scattering relative to that of Raman scattering by a factor of $\frac{2}{3}$. This suppression indicates that the discussion and results from Ref.~\cite{ozeri2007} are not straightforwardly extended to include Rayleigh scattering. We also note that this fidelity is independent of gate duration. The time independence results from the fact that the error is proportional to $\Gamma\tau_g$ and $\Gamma\propto g_0^2$ while, from Eq.~\ref{eq:GeoPhase}, $\tau_g\propto 1/g_0^2$.

We can now calculate an approximate expected error due to spontaneous emission for this gate at our current laser frequency detuning, polarization angle, and center-of-mass mode frequency, where we have $\Delta/(2\pi) = -5.29$~GHz, $\phi_P = \pm 65.3^\circ$, and $\omega_z/(2\pi) = 1.59$~MHz. We also assume a gate comprising a single loop in phase space. To compare the infidelities due to spontaneous emission of gates at high fields with those performed in RF Paul traps, we assume a more similar laser beam geometry to those systems. Leading trapped ion platforms often use angles between their beams of either $90^\circ$ \cite{ballance2016, gaebler2016} or $180^\circ$ \cite{landsman2019} as opposed to the $20^\circ$ currently used in our experiment. These higher angles result in a larger value of $\partial k$ with laser parameters that are otherwise constant, and therefore a higher interaction strength with less spontaneous emission. With an angle between our beams of $\theta_R = 90^\circ$ but the same laser frequency detuning and polarization, we obtain a two-qubit gate error of $8.6\times10^{-3}$, a factor of four lower than with our current separation angle ($\theta_R = 20^\circ$). In the highest fidelity trapped ion gates using ground state qubits and electric dipole transitions, the errors due to spontaneous emission are given as $4\times10^{-4}$ \cite{ballance2016} and $5.7\times10^{-4}$ \cite{gaebler2016}, about an order of magnitude lower than we could achieve with our current laser detuning. These gates, however, were operated with much larger detunings (-1~THz and -730~GHz, respectively) from atomic transitions and therefore required much higher laser intensities.

These results also confirm the discussion in Sec.~\ref{sec:FOM} on the applicability of the linear and quadratic figures of merit. We emphasize that the case considered in this section requires the use of the linear figure of merit, since we are assuming a fixed relationship between $J$, $\tau_g$, and $\delta_z$. The results for the linear figure of merit with the ACSS nulled (Fig.~\hyperref[fig:FOMZZDeltaACSS=0]{\ref*{fig:FOMZZDeltaACSS=0}(b)}) suggest that a two-qubit gate could be significantly improved by operating at a detuning of approximately $300$~GHz. Indeed, at this point, with a polarization $\phi_P = 70.5$ degrees, we compute a gate error of $3.1\times10^{-3}$ for $\theta_R=90^\circ$. The error does not improve by as large of a factor as $\zeta_{L,LS}$ because of the relative suppression of the effect of Rayleigh scattering. If the ACSS is not nulled, large detunings outside of the $P_{1/2}$ and $P_{3/2}$ manifolds would also be a promising option (Fig.~\hyperref[fig:FOMZZVandH]{\ref*{fig:FOMZZVandH}(b)}) if laser power is not a limitation. We discuss these trade-offs further in Sec.~\ref{sec:OperatingPoints}. 

\section{Comparison of figures of merit and required optical powers for potential operating points}\label{sec:OperatingPoints}

\begin{figure*}[t]
    \centering
    \includegraphics[width=\textwidth]{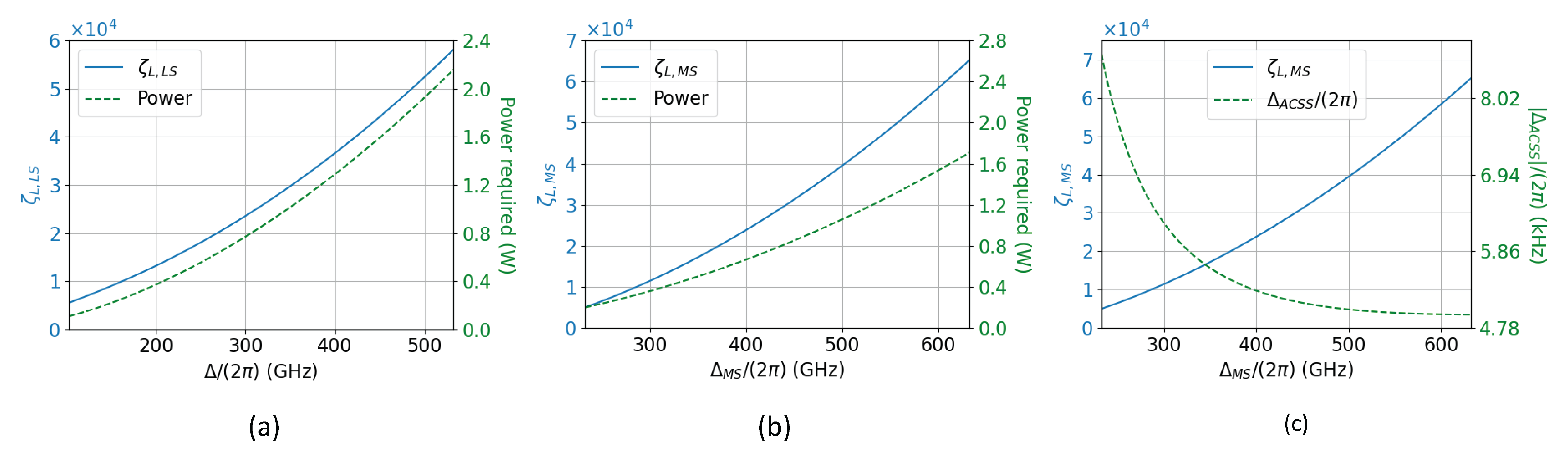}
    \caption{Plots characterizing gate performance at large detunings. (a) $\zeta_{L, LS}$ (blue solid lines) and total power required for a 1~ms two-qubit cat state time with 1~mm beam waists (green dashed lines) versus detuning for a LS gate with vertical polarization. $\Delta_{ACSS}/(2\pi) = -5.4$~kHz for all detunings $\Delta$. (b) $\zeta_{L, MS}$ (blue solid line) and power required for a 1~ms two-qubit cat state time with 1~mm beam waists (green dashed line) versus detuning for a MS gate. We assume the high-frequency beam has $\sigma$ polarization and the lower-frequency beam has $\pi$ polarization and $b_\sigma = b_\pi$. (c) $\zeta_{L,MS}$ (blue solid line) and $\Delta_{ACSS}/(2\pi)$ in kHz (green dashed line) versus detuning for a MS gate with the same configuration as in (b).}
    \label{fig:FarDetuningPlots}
\end{figure*}

In Secs.~\ref{sec:LSSpontaneousEmission} and \ref{sec:MSFOM} we demonstrated that there are a variety of laser detunings and polarizations that produce advantageous figures of merit. However, there are other considerations for implementing these settings in the laboratory. In this section, we select a number of potential operating points and compare them with regards not only to spontaneous emission, but also with regards to the laser power required and the differential AC Stark shift. 
\begin{table*}
    \renewcommand{\arraystretch}{1.5}
    \centering
    \begin{tabular}{|L|N|M|P|P|L|}
    \hline
    \multicolumn{1}{|L|}{\textbf{Operating point number}} & \multicolumn{1}{|N|}{\textbf{Gate type}\vspace{2pt}} & \multicolumn{1}{|M|}{\textbf{Beam detuning and polarization}\vspace{2pt}}& \multicolumn{1}{|P|}{\textbf{Linear figure of merit value}\vspace{2pt}} & \multicolumn{1}{|P|}{\textbf{Power required (mW)}\vspace{2pt}} &
    \multicolumn{1}{|L|}{\textbf{Figure}\vspace{3pt}}\\
    \hline
    1\vspace{1pt} & LS\vspace{1pt} & $\Delta/(2\pi) = -5.29$~GHz, $\phi_P = \pm 65.3^\circ$ & 1066.9\vspace{1pt} & 13.6\vspace{1pt} & Fig.~\hyperref[fig:FOMZZDeltaACSS=0]{\ref*{fig:FOMZZDeltaACSS=0}(b)}\vspace{1pt}\\
    \hline
    2\vspace{1pt} & LS\vspace{1pt} & $\Delta/(2\pi) = 299.3$~GHz, $\phi_P = \pm 70.5^\circ$ & 3440.9\vspace{1pt} & 3443.6 \vspace{1pt} & Fig.~\hyperref[fig:FOMZZDeltaACSS=0]{\ref*{fig:FOMZZDeltaACSS=0}(b)}\vspace{1pt}\\
    \hline
    3\vspace{5pt} & MS\vspace{5pt} & $\Delta_{MS}/(2\pi) = -163.2$~GHz, perpendicular polarizations, $b_\sigma = \sqrt{2}b_\pi$ & 1769.7\vspace{5pt} & 48.9 \vspace{5pt} & Fig.~\hyperref[fig:MSFOM]{\ref*{fig:MSFOM}(b)}\vspace{5pt}\\
    \hline
    4\vspace{1pt} & MS\vspace{1pt} & $\Delta_{MS}/(2\pi) = -142.6$~GHz, $\phi_{SB} = 39.3^\circ \phi_C = -41.5^\circ$ & 1041.2\vspace{1pt} & 20.1 \vspace{1pt} & Fig.~\hyperref[fig:MSFOMnull]{\ref*{fig:MSFOMnull}(b)}\vspace{1pt}\\
    \hline
    5\vspace{1pt} & MS\vspace{1pt} & $\Delta_{MS}/(2\pi) = 431.4$~GHz, $\phi_{SB} = 83.5^\circ, \phi_C = 71.0^\circ$ & 9512.9\vspace{1pt} & 1997.9 \vspace{1pt} & Fig.~\hyperref[fig:MSFOMnull]{\ref*{fig:MSFOMnull}(b)}\vspace{1pt}\\
    \hline
    \end{tabular}
\caption{Various operating points of interest for the gate regime. The points for the LS gate are characterized by a detuning $\Delta$ and a polarization angle $\phi_P$. For the MS gate, we consider both configurations 1 and 2 (see Table~\ref{tab:MS_config}). For configuration 1, there are two relevant parameters that can vary\,---\,the detuning $\Delta$ and the ratio of the electric fields $b_\sigma/b_\pi$. For configuration 2, there are three relevant parameters\,---\,the detuning $\Delta$, the polarization angle for the sideband beam, $\phi_{SB}$, and the polarization angle for the carrier beam, $\phi_C$. $\phi_C$ can be either positive or negative. For each point, we include the value of the linear figure of merit ($\zeta_{L,LS}$ or $\zeta_{L, MS}$) and the total laser power required to achieve a two-qubit cat state time of 1 ms with Gaussian laser beams with a $1/e^2$ waist of 1 mm. We note that operating point 1 corresponds to the current NIST configuration.}
    \label{tab:OperatingPoints}
\end{table*}

For this comparison, we consider the linear figures of merit, $\zeta_{L, LS}$ and $\zeta_{L, MS}$, for the LS and MS gates respectively. We focus here on the gate regime where the spin-dependent optical dipole force frequency is tuned close to the frequency of an ion crystal motional mode. This includes recent experiments in the NIST Penning laboratory where the spin-dependent force frequency is tuned close to the COM mode \cite{garttner2017}. 

In the gate regime, the spin-dependent force, and therefore the laser intensity, required to generate a given entangled spin state is inversely proportional to the gate time $\tau$. In trapped ion experiments in RF Paul traps, laser beams are often focused to have very small waists at the ion locations, especially for gates between pairs of individual ions \cite{Debnath2016}. In the NIST Penning trap, however, in part because the ion crystal is rotating, we implement a global entangling gate and address the whole crystal of ions simultaneously with a nearly uniform electric field. For approximately 100 ions, a uniform laser intensity with a circular beam requires a Gaussian beam waist of $\sim 1$~mm at the plane of the ions. In comparison to the $\sim 10$~$\mu$m waists employed with smaller ion crystals in RF traps, this constraint results in a lower intensity for a given power. Therefore, for Penning traps and large ion crystals in general it is important to consider the power requirements for driving global quantum operations. For the sake of comparison, we choose to consider the power required to achieve a two-qubit spin cat state (or Bell state) in a time $\tau = \pi/(2J)$ \cite{garttner2017} of 1~ms ($\delta_z/(2\pi) = 1$~kHz). We assume that the laser beam is Gaussian and the radius where the laser intensity is reduced by $1/e^2$ is 1~mm. Additional assumptions include a COM mode frequency $\omega_z/(2\pi) = 1.59$~MHz and a 1D difference wave vector $\partial k$ from the use of 313 nm laser beams ($^9$Be$^+$) crossing at $\pm 10^\circ$ as shown in Figs.~\hyperref[fig:BeLevelsODFBeams]{\ref*{fig:BeLevelsODFBeams}(b)} and \ref{fig:MS_Lasers_Penning}.
These parameters result in a single-ion Lamb-Dicke parameter $\eta = \partial k z_0 = 0.13$ where $z_0 = \sqrt{\hbar/(2m\omega_z)}$. As an aside, we note that the cat state time scales with the number of ions $\mathcal{N}$, so for the same laser power and a 100 ion crystal, this is only $2\%$ of the time required to achieve a spin cat state. Previous work at NIST demonstrated coherent quantum evolution with 100 ions to approximately 7$\%$ of the spin cat state time in $\sim 1$~ms \cite{garttner2017}.

Table~\ref{tab:OperatingPoints} lists five operating points, chosen because of their relatively high figure of merit values, that we will examine more closely. For all of these points the AC Stark shift can be nulled. Two of these points are for the LS gate and three are for the MS gate. With points 1, 3, and 4, the optical dipole force lasers are tuned between the Zeeman levels of the excited $P$ state manifolds. With points 2 and 5, the optical dipole force lasers are tuned to the high frequency side of the excited $P$ state manifolds. For the LS gate, an examination of Fig.~\hyperref[fig:FOMZZDeltaACSS=0]{\ref*{fig:FOMZZDeltaACSS=0}(b)} reveals that, other than the current NIST operating point (point 1), there are no operating configurations for which the AC Stark shift can be nulled with a significantly higher values of $\zeta_{L,LS}$ except for point 2 (and other points near it). While point 2 initially seems like a promising option since it has both a high value of $\zeta_{L,LS}$ and no differential ACSS, it requires a laser power that is not readily attainable with current technology. This high power requirement results from both the large detuning and the fact that $F_\uparrow \sim F_\downarrow$ for the polarizations required to null the differential ACSS. The current NIST operating point (point 1 in Table~\ref{tab:OperatingPoints}) is therefore roughly optimal for an LS gate obtained by tuning between the Zeeman levels and nulling the AC Stark shift. Below we will contrast the merits of point 1 with the range of large detunings shown in Figs.~\hyperref[fig:FOMZZVandH]{\ref*{fig:FOMZZVandH}(b)} and Fig.~\hyperref[fig:FarDetuningPlots]{\ref*{fig:FarDetuningPlots}(a)} where much larger figures of merit can be achieved, but with the drawback of a nonzero differential ACSS.

An inspection of Figs.~\hyperref[fig:MSFOM]{\ref*{fig:MSFOM}(b)} and \hyperref[fig:MSFOMnull]{\ref*{fig:MSFOMnull}(b)} indicates that points 3-5 provide some of the best options for an MS gate where the AC Stark shift is nulled. As shown in Fig.~\hyperref[fig:MSFOM]{\ref*{fig:MSFOM}(b)}, point 3 clearly has the highest value of $\zeta_{L,MS}$ of any point in the perpendicular polarization configuration with $\Delta_{ACSS}=0$.\footnote{Point 3 is very close to an atomic resonance ($< 2$~GHz) that can be driven by $\pi$ polarized light in the high frequency beam or $\sigma$ polarized light in the lower frequency beam. Therefore, small polarization impurities result in rapid decreases in the figure of merit as the deviation from the ideal polarization increases. Other potential operating points with similar figures of merit and a nulled ACSS could be obtained by adjusting the ratio $b_\sigma/b_\pi$ to shift the null point away from this resonance.} However, $\Delta_{ACSS}=0$ only when both laser beams are applied to the ions, which makes optimizing the gate more technically challenging and increases the sensitivity to power fluctuations that change the relative intensity of the two beams. 

From Fig.~\hyperref[fig:MSFOMnull]{\ref*{fig:MSFOMnull}(b)}, the point with the highest value of $\zeta_{L,MS}$ that is not far-detuned from resonance is point 4. This point addresses the issue with the differential ACSS since it is nulled with either beam individually and we would therefore be insensitive to power fluctuations in either beam, even if they are not common-mode. The value $\zeta_{L, MS}$ is somewhat lower than that for point 3, although still comparable to the value of $\zeta_{L, LS}$ for point 1. Finally, point 5 is similar to point 2 in that it has a low impact from spontaneous emission and results in $\Delta_{ACSS}=0$ with either beam, but requires higher laser power than is readily obtained in the laboratory. 

For both the LS and MS gates, the linear figure of merit rapidly increases for large detunings outside the excited $P$ state manifolds. At low fields, gates have been performed with detunings of tens of THz \cite{hayes2012}, but with much more tightly focused laser beams \cite{choi2014}. However, as discussed above, global entangling gates on large ion crystals require a much larger beam waist and therefore higher power to achieve a given intensity. Thus, the discussion here focuses on the ability to optimize simultaneously for both coherence and lower laser power.

We are interested in knowing how large a linear figure of merit we could achieve with larger detunings and a given laser power. In Figs.~\hyperref[fig:FarDetuningPlots]{\ref*{fig:FarDetuningPlots}(a)} and \hyperref[fig:FarDetuningPlots]{(b)} we plot the figure of merit and power required versus detuning for the LS and MS gates respectively. For the LS gate, we assume vertical ($\pi$) polarization in both beams, and for the MS gate we assume the perpendicular polarization configuration (configuration 1) with equal intensities. We note that for either gate, with a power of about 250-350 mW total we can attain figure of merit values of at least 10000, which will reduce errors due to spontaneous emission by about an order of magnitude compared to the current operating point (point 1).

\begin{figure*}[t]
    \centering
    \includegraphics[width=\textwidth]{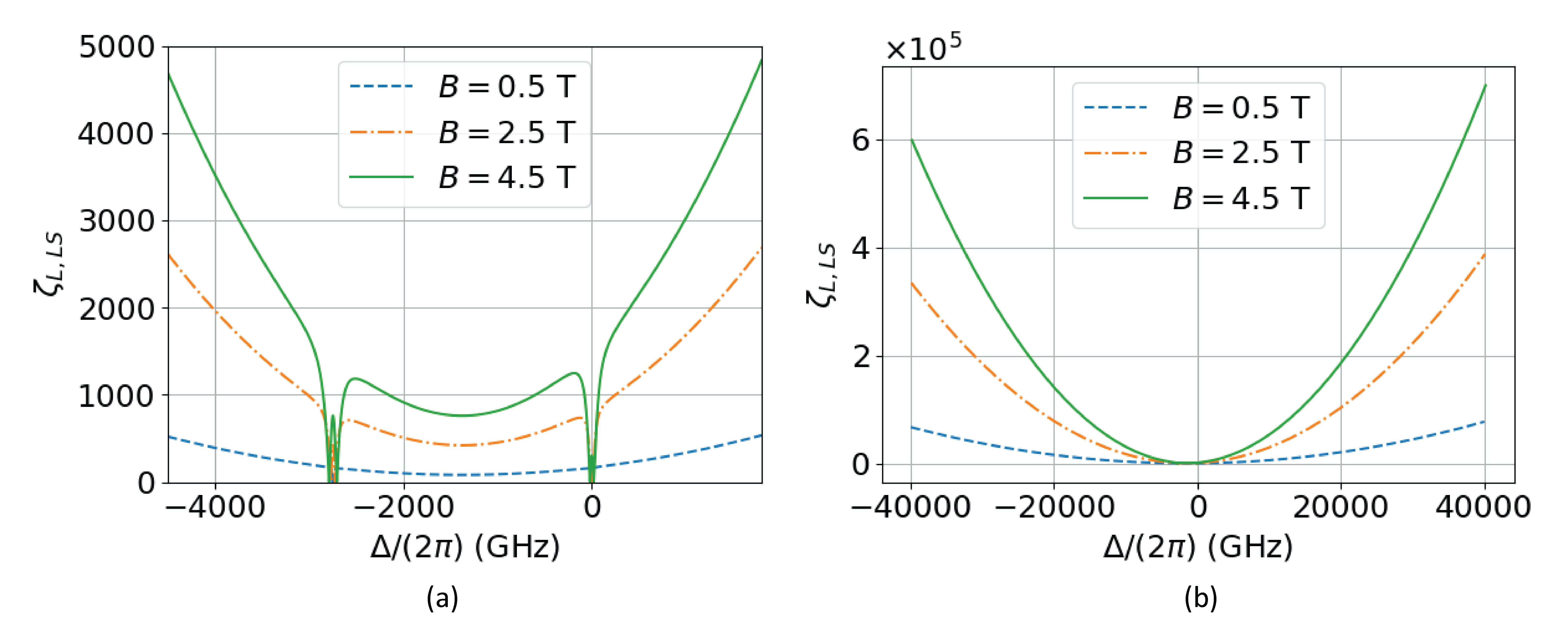}
    \caption{Plots of the linear figure of merit for an LS gate with $^{24}$Mg$^+$ for magnetic fields of 0.5~T (blue dashed line), 2.5~T (orange dashed and dotted line), and 4.5~T (solid green line). Vertical polarization is assumed, although the results are similar for horizontal polarization as well. These plots can be compared with those for $^9$Be$^+$ in Figs.~\hyperref[fig:FOMZZlinmultiB]{\ref*{fig:FOMZZlinmultiB}(b)} and \hyperref[fig:FOMZZlinmultiB]{(c)}. (a) $\zeta_{L,LS}$ vs $\Delta$ for $^{24}$Mg$^+$. The zero field frequency difference between $S_{1/2}$ and $P_{1/2}$ corresponds to a detuning equal to the fine structure splitting, approximately 2750~GHz. (b) $\zeta_{L,LS}$ vs $\Delta$ for $^{24}$Mg$^+$ with large values of $\Delta$ compared to the fine structure splitting. For a given value of $\Delta$, $\zeta_{L,LS}$ is approximately 300 times smaller for Mg$^+$ than for Be$^+$. (Note the difference in the powers of 10 on the $y$ axis.)}
    \label{fig:MgPlotsLS}
\end{figure*}

There are two disadvantages to the far-detuned configuration, however. First, the laser power required to achieve an order of magnitude increase in $\zeta_{L,LS}$ or $\zeta_{L,MS}$ is, in general, more than an order of magnitude higher than that required for tuning between the Zeeman levels (i.e. points 1 and 4). Even 300~mW, while commercially available, can be a technically challenging amount of power to maintain at UV wavelengths (313~nm for $^9$Be$^+$). We note, though, that elliptical laser beams with a reduced vertical waist can be utilized for the setup shown in Figs.~\hyperref[fig:BeLevelsODFBeams]{\ref*{fig:BeLevelsODFBeams}(b)} and \ref{fig:MS_Lasers_Penning} and currently employed at NIST. If the optical dipole force laser beams cross the ion crystal at a $10^\circ$ angle, the vertical waist of the laser beam can be reduced by about a factor of 6, resulting in the same reduction in the required laser power. 

A second disadvantage is that the differential AC Stark shift is not nulled for large detunings, so experiments would be more sensitive to intensity fluctuations. With pure vertical (horizontal) polarization, both $F_0$ and $\Delta_{ACSS}$ are proportional to $A(B)_\uparrow - A(B)_\downarrow$ (see Eqs.~\ref{eq:ACSS_Individual}, \ref{eq:ACSS_diff}, and \ref{eq:F_0updown}). Since $F_0$ is fixed by the desired gate time and geometric phase accumulation, $\Delta_{ACSS}$ is as well. For the parameters used for the plot in Fig.~\hyperref[fig:FarDetuningPlots]{\ref*{fig:FarDetuningPlots}(a)}, $\Delta_{ACSS}/(2\pi) = -5.4$~kHz. The differential ACSS varies with detuning for the MS gate, so we plot $\Delta_{ACSS}$ along with $\zeta_{L,MS}$ versus detuning in Fig.~\hyperref[fig:FarDetuningPlots]{\ref*{fig:FarDetuningPlots}(c)}. The absolute value of $\Delta_{ACSS}/(2\pi)$ remains in the range of $\lesssim 8$~kHz and decreases to a nearly constant value of $\sim 5$~kHz as $\Delta_{MS}$ increases.

For the LS gate, we can use a spin-echo sequence to cancel the impact of the differential ACSS as long as the intensity in the laser beams remains stable between the two arms of the spin-echo sequence. An intensity imbalance between the two arms will produce a rotation about the $z$-axis of the Bloch sphere. As a very rough upper bound on the required stability, we estimate the intensity imbalance that will produce a rotation comparable to the angle defined by projection noise ($\sqrt{\mathcal{N}}/2$) and the length of the composite Bloch vector ($\mathcal{N}/2$). For $\mathcal{N}=100$ this corresponds to an angle of $1/\sqrt{\mathcal{N}} \approx 0.1$ rad. For the generation of spin-squeezed states, for example, the stability should produce rotations significantly less than this. With a 1~ms arm time, appropriate for generating a spin-squeezed state with $\mathcal{N}=100$ spins and a $\sim$5~kHz AC Stark shift, the Bloch vector will undergo a rotation of $10\pi$ rad during a single arm time. To cancel out this rotation to better than 0.1 rad requires a laser intensity stability between the two arms of the spin-echo sequence of better than $0.1/(10\pi)\sim 3\times10^{-3}$. More highly entangled states can require longer interaction times and be more sensitive to erroneous rotations, dictating even more stringent intensity stabilities than $10^{-3}$, which can be technically challenging to achieve. An alternative calculation of this requirement for an LS gate, albeit with different experimental parameters, estimates a required intensity stability of $10^{-4}$ and states this is potentially feasible on the timescale of a typical spin-echo sequence \cite{sawyer2021,thom2018}. We note that due to the spatial extent of the ion crystal the AC Stark shift could be inhomogeneous across the crystal. For a 200 ion crystal, which has a radius of order 0.1~mm, magnetic field inhomogeneities will be very small ($<10^{-8}$~T). Therefore, we expect the dominant spatial inhomogeneity in the AC Stark shift will be due to variation in the laser intensity across the crystal.

\section{Fine Structure Dependence of the Figures of Merit}\label{sec:MgComparison}

To illustrate the generality of the calculations presented in this paper and how the details change with atomic properties, we calculate here the equivalent figures of merit for LS and MS gates performed in $^{24}$Mg$^+$, the second lightest alkaline earth element after $^9$Be$^+$ and another ion frequently employed in trapped-ion quantum processing. The only significant differences in the atomic structure between $^{24}$Mg$^+$ and $^9$Be$^+$ at a high magnetic field are the much larger fine structure splitting $\omega_{FS}$ between the $P_{1/2}$ and $P_{3/2}$ levels\,---\,$2745$~GHz  and $197$~GHz respectively\,---\,and the larger decay rate $\gamma$ from the excited states\,---\,$2\pi\times41$~MHz versus $2\pi\times17$~MHz respectively. Na\"{i}vely, one might expect a decrease in the figure of merit by a factor of the ratio of the decay rates, which is roughly 2.5. However, the results actually depend on the fine structure splitting as well.

\begin{figure*}[t]
    \centering
    \includegraphics[width=\textwidth]{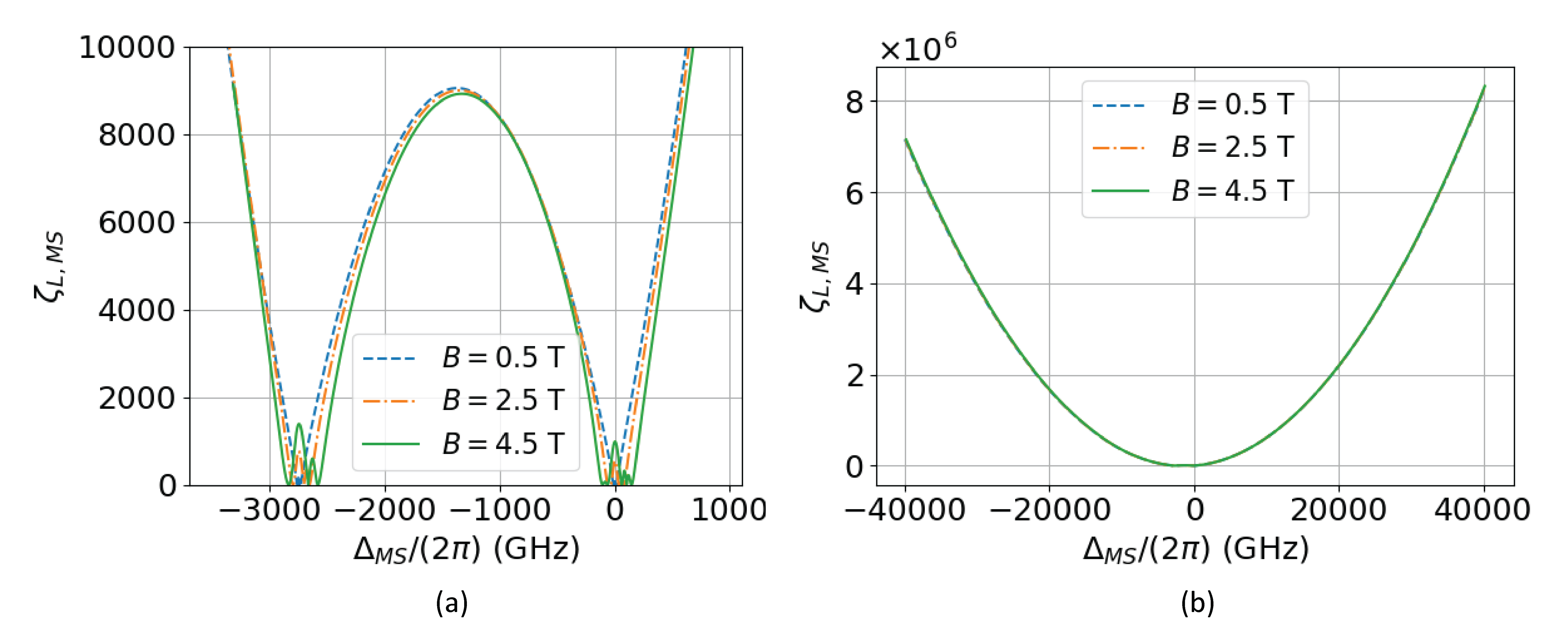}
    \caption{Plots of the linear figure of merit for an MS gate with $^{24}$Mg$^+$ for magnetic fields of 0.5~T (blue dashed line), 2.5~T (orange dashed and dotted line), and 4.5~T (solid green line). These results are for configuration 1 (see Table~\ref{tab:MS_config}) and can be compared with the plots for $^9$Be$^+$ in Fig.~\ref{fig:MSmultiB}. (a) $\zeta_{L,MS}$ vs $\Delta_{MS}$ for $^{24}$Mg$^+$. (b) $\zeta_{L,MS}$ vs $\Delta_{MS}$ for large detunings. For a given detuning far from any resonance, $\zeta_{L,MS}$ is approximately 30 times lower for Mg$^+$ than for Be$^+$.}
    \label{fig:MgPlotsMS}
\end{figure*}

The results for the LS gate in $^{24}$Mg$^+$ are shown in Fig.~\ref{fig:MgPlotsLS} and can be compared with the results for $^9$Be$^+$ in Fig.~\ref{fig:FOMZZlinmultiB}. In Fig.~\hyperref[fig:MgPlotsLS]{\ref*{fig:MgPlotsLS}(a)}, the sharp decreases in the figure of merit near 0~GHz and -2750~GHz correspond to tuning within the $P_{3/2}$ Zeeman levels and tuning between the $P_{1/2}$ Zeeman levels respectively. In these regions, $\zeta_{L,LS}$ is suppressed relative to the value for Be$^+$, but only by a factor comparable to the ratio of the decay rates. We also see that Mg$^+$ allows for operating between the $P_{1/2}$ and $P_{3/2}$ manifolds with large detunings, although the values of the figure of merit are not significantly larger than that for the current NIST operating point. On the other hand, a comparison of the plot in Fig.~\hyperref[fig:MgPlotsMS]{\ref*{fig:MgPlotsLS}(b)} with Fig.~\hyperref[fig:FOMZZlinmultiB]{\ref*{fig:FOMZZlinmultiB}(c)} shows that, for large detunings $\Delta$, $\zeta_{L,LS}$ is decreased by a factor of about 300 for Mg$^+$ relative to the corresponding value for Be$^+$ for large detunings. The strong dependence on the fine structure splitting $\omega_{FS}$ is in agreement with the analytical treatment of Appendix~\ref{appendix:BomegaFS_dependence}, where for large detunings $\Delta$ it is shown that
\begin{equation}
    \zeta_{L,LS}\sim\left(\frac{\Delta_z}{\omega_{FS}}\right)\left(\frac{\Delta^2}{\gamma\omega_{FS}}\right).
    \label{eq:FOMLSScaling}
\end{equation}
The $1/\omega_{FS}^2$ scaling results from the decoherence due to Raman scattering. 

The MS gate suffers less from the increase in fine structure splitting, as demonstrated by a comparison of Fig.~\ref{fig:MgPlotsMS} with Fig.~\ref{fig:MSmultiB}. An interesting feature of the MS linear figure of merit is the very high value one would obtain with a virtual level approximately halfway between the $P_{1/2}$ and $P_{3/2}$ manifolds in ions with larger fine structure splittings than $^9$Be$^+$. For large detunings outside of the excited state manifolds, however, $\zeta_{L,MS}$ is suppressed by a factor of approximately 30 for Mg$^+$ compared to Be$^+$. This result is consistent with the results from Appendix~\ref{appendix:BomegaFS_dependence} that show
\begin{equation}
    \zeta_{L, MS} \sim \frac{\Delta_{MS}^2}{\gamma\omega_{FS}}.
    \label{eq:FOMMSScaling}
\end{equation}
The scaling of the decoherence with the fine structure splitting is the same as for the LS gate ($\propto \omega_{FS}^2$), but $\Omega_R$ scales linearly with the fine structure splitting as well, which results in an overall dependence of the figure of merit that is inversely proportional to the fine structure splitting.

Comparing Eqs.~\ref{eq:FOMLSScaling} and \ref{eq:FOMMSScaling}, we see that the relative performance of the LS and MS gates $\zeta_{L,LS}/\zeta_{L,MS}$ scales as $\Delta_z/\omega_{FS}$ (treating the detunings for the two gates as equivalent). Thus, for lower magnetic fields and larger fine structure splittings, the MS gate will be preferable. The strong dependence on the fine structure for the LS gate is reduced when $\Delta_z\gtrsim\omega_{FS}$. For any ion heavier than $^9$Be$^+$, however, this condition requires magnetic fields at least an order of magnitude larger than the 4.5~T field currently employed at NIST. Because such magnets do not exist, the use of an LS gate in an ion with a larger fine structure splitting requires increasing the angle between the beams so that the strengths of transitions driven by $\sigma^-$ polarized light can differ from the strengths of transitions driven by $\sigma^+$ polarized light. This would enable a different force on the two qubit states that arises not only from differences in detuning but also in the coupling to different excited states.

Additional modifications for the rates of decoherence due to Raman scattering would be necessary to extend this analysis to account for low-lying $D$ manifolds in heavier ions, but will rely on the same principles discussed in this paper. This analysis shows, however, not only that the methods demonstrated in this paper can be extended to other ions, but also that for the configurations described here, ions with smaller fine structure splittings are preferable, especially for the LS gate.

\section{Conclusion}\label{sec:Conclusion}

We have presented a thorough exploration of the interaction strengths and spontaneous emission errors obtained with different laser detunings and polarizations for multiqubit gates at high magnetic fields. We focused on laser beam configurations illustrated in Figs.~\ref{fig:BeLevelsODFBeams} and \ref{fig:MS_Lasers_Penning} where the optical dipole force laser beams have waists that are large compared to the size of the ion crystal and are directed nearly perpendicular to the magnetic field axis. This geometry is suitable for interacting with large, rotating ion crystals where it is necessary to align the difference wavevector $\partial\vec{k}$ of the laser beams parallel to the magnetic field and rotation axis of the crystal. The calculations are especially relevant for Penning traps, where, due to the high magnetic fields, the Zeeman splittings can be comparable to the fine structure, and where the large ion crystals result in tight constraints on laser power. 

We also discussed the trade-offs between tuning the optical dipole force laser beams between the Zeeman levels, enabling nulling of the AC Stark shift and operation at lower laser powers, versus tuning outside the Zeeman levels where the impact of off-resonant light scattering can be mitigated at the expense of larger required laser power. While the exact detunings and polarizations of the operating points discussed here are specific to $^9$Be$^+$ and will vary depending on the magnetic field and atomic properties, the methodology used to obtain optimal operating points and compare them remains generally valid for any spin-zero nucleus or nonzero spin nucleus that can be optically pumped to a single nuclear spin level such as $^9$Be$^+$.

Furthermore, we have concluded that the M\o lmer-S\o rensen gate can be performed in a Penning trap with similar performance with regards to spontaneous emission as the light-shift gate, expanding the types of simulations and measurements that can be performed in Penning traps. We have also shown, for laser beam configurations where the beams are directed approximately perpendicular to the magnetic field (see Figs.~\ref{fig:BeLevelsODFBeams} and \ref{fig:MS_Lasers_Penning}), that the M\o lmer-S\o rensen gate is preferable to the light-shift gate when Zeeman splittings are small compared to the fine structure splitting ($\Delta_z/\omega_{FS} \ll 1$). Finally, we have shown that errors due to off-resonant scattering in the LS gate currently employed at NIST (operating point 1 in Table~\ref{tab:OperatingPoints}) are larger than in state-of-the-art two-qubit gates performed at low magnetic field, but that there are a number of promising operating points that can reduce these errors by an order of magnitude.

%\section*{Acknowledgments}
\begin{acknowledgments}
We thank J. Stuart and B. Sawyer for comments on the manuscript.  This work is supported by a collaboration between the U.S. Department of Energy, Office of Science and other agencies.  This material is based upon work supported by the U.S. Department of Energy, Office of Science, NQI Science Research Centers, Quantum Systems Accelerator (QSA). A.M.R. acknowledges additional support from NSF PHY 1820885 and NSF PFC PHY- 1734006. J.J.B. acknowledges additional support from the DARPA ONISQ program and AFOSR grant FA9550-201-0019.
\end{acknowledgments}
\appendix
\section{Calculation of the excited state Zeeman splittings and matrix elements at high fields}\label{sec:AppEnergies}
See Ref.~\onlinecite{bethe1957}, Secs. 45-47 for a more detailed treatment of the Zeeman effect for $^2P$ levels. For $^9$Be$^+$ and the 4.5 T magnetic field, the fine structure splitting between the $^2P_{1/2}$ and $^2P_{3/2}$ levels is a similar order of magnitude to the Zeeman splitting. As a result, we cannot treat either the magnetic field Hamiltonian or the fine structure Hamiltonian as a perturbation and must diagonalize the full Hamiltonian. For the numerical calculations presented in this paper, we find the energy levels using the full diagonalization of the Hamiltonian, but in calculating matrix elements (for example, Eqs. (4) and (9)) relevant for AC Stark shifts and off-resonant light scattering we do not account for the small admixture of the  $\Ket{^2P_{1/2},m_J =+1/2}$ and $\Ket{^2P_{3/2},m_J =+1/2}$ levels or the $\Ket{^2P_{1/2},m_J=-1/2}$ and $\Ket{^2P_{3/2}, m_J =-1/2}$ levels. In other words, we assume that $J$ is still a good quantum number, although this becomes progressively less accurate as we increase the magnetic field strength. For a field of 4.46~T, we numerically estimate that the errors from this approximation are limited to $\lesssim 10\%$. If significantly higher fields were to be considered using the methodology employed in this paper, however, it would be necessary to account for this effect. Because of this approximation and because most experiments with trapped ions operate at fields that are $\lesssim 4.5$~T, we choose to set this magnetic field as the maximum we consider.

\section{Calculation of the decoherence rates}\label{sec:AppA}

Here we calculate the appropriate single-spin decoherence rate to use for the LS and MS interactions. For both the LS and MS gates, we use the Lindblad master equation,
\begin{equation}
    \frac{d\hat{\rho}}{dt} = -\frac{i}{\hbar}\left[\hat{H}_1, \hat{\rho}\right] + \sum_j\mathcal{L}\left[\hat{O}_j\right]\hat{\rho},
    \label{eq:MasterEq}
\end{equation}
to find the time evolution of the density matrix of a single-ion qubit. In this equation, $\hat{H}_1$ is an effective single ion qubit Hamiltonian with either the LS or MS gate interaction, $\hat{O}_k$ are the jump operators that encapsulate the decoherence due to spontaneous emission, and $\mathcal{L}\left[\hat{O}_j\right]\hat{\rho}$ is given by
\begin{equation}
    \mathcal{L}\left[\hat{O}_j\right]\hat{\rho} = \hat{O}_j\hat{\rho}\hat{O}_j^\dagger - \frac{1}{2}\left(\hat{O}_j^\dagger\hat{O}_j\hat{\rho} + \hat{\rho}\hat{O}_j^\dagger\hat{O}_j\right).
    \label{eq:Lindbladian}
\end{equation}

Equation~\ref{eq:Ising_general} described the interaction between the ions engineered by the LS or MS laser beams. To consider the effect on a single ion within a multi-ion crystal, we use the mean-field approximation for both the LS and MS gates. The single-particle Hamiltonian for ion $i$ can then be written as \cite{Britton2012}
\begin{equation}
    \hat{H}_1 = \hbar \bar{B}_i \hat{\sigma}_i^\alpha,
    \label{eq:H_singleion}
\end{equation}
where
\begin{equation}
    \bar{B}_i = \frac{1}{\mathcal{N}}\sum_{j\neq i}^{\mathcal{N}-1} J_{ij} \left\langle \hat{\sigma}_j^\alpha \right\rangle,
    \label{eq:chi}
\end{equation}
and $\alpha$ indicates $z$ or $x$ for the LS or MS gates, respectively. For simplicity, we will drop the subscript $i$ for the remainder of the appendix.

Next, we consider the jump operators $\hat{O}_k$. For the LS gate, these are given by \cite{fossfeig2013}
\begin{align}
    \hat{O}_{\uparrow\downarrow} &= \sqrt{\Gamma_{\uparrow\downarrow}}\hat{\sigma}^-, \nonumber\\
    \hat{O}_{\downarrow\uparrow} &= \sqrt{\Gamma_{\downarrow\uparrow}}\hat{\sigma}^+, \nonumber\\
    \hat{O}_{el} &= \sqrt{\frac{\Gamma_{el}}{4}} \hat{\sigma}^z.
    \label{eq:LS_jump}
\end{align}
We note that there is a factor of $\sqrt{2}$ discrepancy in these definitions from those in Ref.~\onlinecite{fossfeig2013}, which arises from writing the master equation slightly differently. The rates $\Gamma_{\uparrow\downarrow}$, $\Gamma_{\downarrow,\uparrow}$, and $\Gamma_{el}$ here are the two-beam rates given in Eqs.~\ref{eq:Gamma_ij} and \ref{eq:Gamma_el}.

We next combine Eqs.~\ref{eq:MasterEq}, \ref{eq:H_singleion}, and \ref{eq:LS_jump} to find a differential equation for the density matrix of a single ion with the LS gate applied. We define the elements of the density matrix as
\begin{equation}
    \hat{\rho} = \begin{pmatrix} \rho_{uu} & \rho_{ud} \\ \rho_{du} & \rho_{dd} \end{pmatrix}.
    \label{eq:rho}
\end{equation}
The time evolution is then 
\begin{widetext}
\begin{equation}
    \frac{d\hat{\rho}}{dt} = \begin{pmatrix} -\Gamma_{\uparrow\downarrow}\rho_{uu} + \Gamma_{\downarrow\uparrow}\rho_{dd} & \left[-2i\bar{B} - \frac{1}{2}\left(\Gamma_{r} + \Gamma_{el}\right)\right]\rho_{ud} \\
    \left[2i \bar{B} - \frac{1}{2}\left(\Gamma_r + \Gamma_{el}\right)\right]\rho_{du} & \Gamma_{\uparrow\downarrow}\rho_{uu} - \Gamma_{\downarrow\uparrow}\rho_{dd} \end{pmatrix}.
\end{equation}
\end{widetext}

The form of this equation is similar to Eq. 8 in Ref.~\onlinecite{Uys2010}, but in the derivation we have included the coherent interaction resulting in the $i\bar{B}$ terms in the off-diagonal matrix elements. The coherent interaction produces simple spin precession about the $z$ axis, and we are interested in the rate at which the spin precession decays. The solutions for the off-diagonal density matrix elements are
\begin{align}
    \rho_{ud}(t) &= \rho_{ud}(0)e^{-2i\bar{B} t}e^{-\frac{1}{2}\left(\Gamma_r + \Gamma_{el}\right)t}, \nonumber\\
    \rho_{du}(t) &= \rho_{du}(0)e^{2i\bar{B} t}e^{-\frac{1}{2}\left(\Gamma_r + \Gamma_{el}\right)t}.
    \label{eq:LS_rho_elements}
\end{align}
It is clear, then, that for the LS gate, the appropriate decoherence rate to use is $\Gamma = \frac{1}{2}\left(\Gamma_r + \Gamma_{el}\right)$ as in Eq.~\ref{eq:Gamma_LS}. The coherent interaction thus does not modify the decoherence rate given in Ref.~\onlinecite{Uys2010}.

For the MS gate, the jump operators have the same form, assuming the red and blue sidebands are detuned by a nonzero amount from resonance. We now define $\Gamma_{\uparrow\downarrow}$, $\Gamma_{\downarrow\uparrow}$, and $\Gamma_{el}$ to be the sum of the rates calculated for the carrier and sideband laser beams. The coherent $\hat{\sigma}^x$ term in the master equation for the MS gate results in spin precession about the $x$ axis. It generates  coupling between different elements of the density matrix. The full differential equation is
\begin{widetext}
\begin{equation}
    \frac{d\hat{\rho}}{dt} = \begin{pmatrix}
    -i\bar{B}\left(\rho_{du}-\rho_{ud}\right) -\Gamma_{\uparrow\downarrow}\rho_{uu} + \Gamma_{\downarrow\uparrow}\rho_{dd} & -i\bar{B}\left(\rho_{dd} - \rho_{uu}\right) - \frac{1}{2}\left(\Gamma_r + \Gamma_{el}\right)\rho_{ud}\\
    i\bar{B}\left(\rho_{dd} - \rho_{uu}\right) - \frac{1}{2}\left(\Gamma_r + \Gamma_{el}\right)\rho_{du} & i\bar{B}\left(\rho_{du} - \rho_{ud}\right)-\Gamma_{\downarrow\uparrow}\rho_{dd} + \Gamma_{\uparrow\downarrow}\rho_{uu}
    \end{pmatrix}.
\end{equation}
\end{widetext}
It is no longer clear from the differential equations alone what the qualitative time dependence of each density matrix element is given the coupling between the elements. This system of equations, along with the condition $\textrm{Tr}\left[\hat{\rho}\right]=1$, can be solved analytically, however. The resulting equations for the time evolution are
\begin{widetext}
\begin{align}
    \rho_{uu}(t) &= \frac{1}{4\bar{B}}\left\{\left[C_2 r - C_3 \left(\Gamma_{el} - \Gamma_r\right)\right]\cos\left(\frac{1}{4}r t\right) + \left[C_2\left(\Gamma_{el} - \Gamma_r\right) + C_3 r\right]\sin\left(\frac{1}{4}rt\right)\right\}e^{-\frac{1}{4}\left(\Gamma_{el} + 3\Gamma_r\right) t} + \kappa_1, \\
    \rho_{ud}(t) &= C_1 e^{-\frac{1}{2}\left(\Gamma_{el} + \Gamma_r\right)t} - 2i\left[C_3 \cos\left(\frac{1}{4}r t\right) - C_2 \sin\left(\frac{1}{4}rt\right)\right]e^{-\frac{1}{4}\left(\Gamma_{el} + 3\Gamma_r\right) t} + \kappa_2,\\
    \rho_{du}(t) &= C_1 e^{-\frac{1}{2}\left(\Gamma_{el} + \Gamma_r\right) t} + 2i\left[C_3 \cos\left(\frac{1}{4}rt\right) - C_2 \sin\left(\frac{1}{4}rt\right)\right]e^{-\frac{1}{4}\left(\Gamma_{el} + 3\Gamma_r\right) t} + \kappa_3,
\end{align}
\end{widetext}
where $C_1$, $C_2$, and $C_3$ are real constants that describe the initial state of the ion, $r = \sqrt{64\bar{B}^2 -\left(\Gamma_{el}-\Gamma_r\right)^2}$ is a modified effective two-photon Rabi frequency, and $\kappa_1$, $\kappa_2$, and $\kappa_3$ are constants describing the state of the ion as $t\rightarrow \infty$. The form of these constants does not impact the results in this paper so we do not include the full expressions. The important feature of these equations is that all oscillatory terms, which describe spin precession, decay with the rate $\Gamma_{MS} = \frac{1}{4}\left(\Gamma_{el} + 3\Gamma_r\right)$ we used in Sec.~\ref{sec:MSFOM}. We have demonstrated here that the inclusion of a coherent interaction can modify the effective decoherence rate due to spontaneous emission, which was not discussed in Ref.~\onlinecite{Uys2010}.

We note also that for the case where one is interested in performing resonant stimulated Raman $\hat{\sigma}^x$ rotations on the qubit (as opposed to driving a full MS interaction), the jump operators are significantly different due to interference between Raman and Rayleigh scattering. While these jump operators will affect other elements of the dynamics of the qubit state, they do not change the rates of decay. We expect that more detail on these effects will be discussed in a future publication.

\section{Magnetic field and fine structure dependence of the linear figures of merit}\label{appendix:BomegaFS_dependence}

We showed with our comparisons of the gates at varying magnetic fields in both $^9$Be$^+$ and $^{24}$Mg$^+$ that the LS and MS gates both depend directly on the fine structure splitting, although the LS gate has a much stronger dependence. The LS gate also strongly depends on the magnetic field. In this appendix, we consider the configurations we used for Figs.~\hyperref[fig:FOMZZlinmultiB]{\ref*{fig:FOMZZlinmultiB}(c)}, \hyperref[fig:MgPlotsLS]{\ref*{fig:MgPlotsLS}(b)}, \hyperref[fig:MSmultiB]{\ref*{fig:MSmultiB}(b)}, and \hyperref[fig:MgPlotsMS]{\ref*{fig:MgPlotsMS}(b)} with large $\Delta$ or $\Delta_{MS}$ and show analytically how both figures of merit depend on the detuning, the magnetic field, and the fine structure splitting.  These results agree qualitatively with the trends observed in the figures listed above.

For these calculations, we define a frequency splitting proportional to the magnetic field $\Delta_z = \frac{1}{2}\Delta_{\uparrow\downarrow} = \frac{\mu_B}{\hbar}B$. For all of these calculations, we will neglect the small hyperfine splittings and  nonlinearities in the splittings between energy levels in the $P$ manifolds (discussed in Appendix~\ref{sec:AppEnergies}).

\subsection{Light-shift gate}
We assume vertically polarized light for both beams and consider detunings $\Delta$ much larger than $\Delta_z$. Ultimately, we will also consider the limit where $\omega_{FS} \ll \Delta$, but to begin, we leave the relative size of the fine structure splitting undetermined. 

First, we consider the spin-dependent force. For vertical polarization,
\begin{equation}
    F_0/(\hbar\partial k) \propto \Delta_{ACSS}.
\end{equation}
Explicitly,
\begin{align}
    F_0/(\hbar \partial k) &\propto \frac{g_0^2}{3}\left[\left(\frac{1}{\Delta + \omega_{FS} + \frac{2}{3}\Delta_z} + \frac{2}{\Delta + \frac{1}{3}\Delta_z}\right)\right.\nonumber\\
    &\left.- \left(\frac{1}{\Delta + \omega_{FS} -\frac{2}{3}\Delta_z} + \frac{2}{\Delta - \frac{1}{3}\Delta_z}\right)\right]
\end{align}
where the first and second terms are the ACSS on $\ket{\uparrow}$ while the third and fourth are the ACSS on $\ket{\downarrow}$. The first and third terms result from coupling to the $P_{1/2}$ manifold, and the second and fourth terms result from coupling to the $P_{3/2}$ manifold.
We factor out $\frac{1}{\Delta}$ and expand to first order in $\frac{\Delta_z}{\Delta}$. The resulting approximate spin-dependent force is
\begin{equation}
    F_0/(\hbar\partial k) \approx \frac{4g_0^2}{9}\frac{\Delta_z}{\Delta^2}\left[1 + \left(\frac{\Delta}{\Delta + \omega_{FS}}\right)^2\right].\label{eq:F0approx}
\end{equation}

Next, for decoherence due to Raman scattering, we have (see Eqs.~\ref{eq:Gamma_ij_singlebeam} and \ref{eq:Gamma_r_ZZ_singlebeam}),
\begin{align}
    \Gamma_{\uparrow\downarrow, 1} &= \frac{2g_0^2\gamma}{9\Delta}\left(\frac{1}{\Delta + \frac{1}{3}\Delta_z} - \frac{1}{\Delta + \omega_{FS} + \frac{2}{3}\Delta_z}\right)^2\nonumber\\
    &\approx \frac{2g_0^2\gamma}{9\Delta^2}\left[\left(\frac{\omega_{FS}}{\Delta + \omega_{FS}}\right)^2 -\frac{2}{3}\frac{\Delta_z}{\Delta}\left(\frac{\omega_{FS}}{\Delta + \omega_{FS}}\right)\right.\nonumber\\
    &\left.\times\left(\frac{\Delta^2 - 2\omega_{FS}\Delta - \omega_{FS}^2}{\left(\Delta+\omega_{FS}\right)^2}\right)\right],
\end{align}

\begin{align}
    \Gamma_{\downarrow\uparrow, 1}&= \frac{2g_0^2\gamma}{9\Delta}\left(\frac{1}{\Delta-\frac{1}{3}\Delta_z} - \frac{1}{\Delta + \omega_{FS} -\frac{2}{3}\Delta_z}\right)^2\nonumber\\
    &\approx \frac{2g_0^2\gamma}{9\Delta^2}\left[\left(\frac{\omega_{FS}}{\Delta + \omega_{FS}}\right)^2 + \frac{2}{3}\frac{\Delta_z}{\Delta}\left(\frac{\omega_{FS}}{\Delta + \omega_{FS}}\right)\right.\nonumber\\
    &\left.\times\left(\frac{\Delta^2 - 2\Delta\omega_{FS} - \omega_{FS}^2}{\left(\Delta+\omega_{FS}\right)^2}\right) \right],\\
    \Rightarrow \Gamma_{r,1} &\approx\frac{4g_0^2\gamma}{9\Delta^2}\left(\frac{\omega_{FS}}{\Delta + \omega_{FS}}\right)^2.
\end{align}
We also explicitly write out the expression for elastic scattering in terms of $\Delta_z$ and $\omega_{FS}$ (Eq.~\ref{eq:Gamma_el_ZZ_singlebeam}),
\begin{align}
    \Gamma_{el, 1} &=\frac{g_0^2\gamma}{9\Delta}\left[\frac{1}{\Delta +\omega_{FS} -\frac{2}{3}\Delta_z} - \frac{1}{\Delta + \omega_{FS} + \frac{2}{3}\Delta_z}\right.\nonumber\\
    &\left. + 2\left(\frac{1}{\Delta-\frac{1}{3}\Delta_z} - \frac{1}{\Delta + \frac{1}{3}\Delta_z}\right)\right]^2\nonumber\\
    &\approx \frac{16g_0^2\gamma}{81\Delta^2}\left(\frac{\Delta_z}{\Delta}\right)^2\left(\frac{2\Delta^2 + 2\omega_{FS}\Delta + \omega_{FS}^2}{\left(\Delta+\omega_{FS}\right)^2}\right)^2.
\end{align}

Since the leading-order term in $\Gamma_{el, 1}$ is $\mathcal{O}\left(\left(\Delta_z/\Delta\right)^2\right)$, when we calculate the overall decoherence rate $\Gamma = \Gamma_{r,1} + \Gamma_{el, 1}$, only $\Gamma_{r,1}$ will contribute significantly, and we find
\begin{equation}
    \Gamma \approx \frac{4g_0^2\gamma}{9\Delta^2}\left(\frac{\omega_{FS}}{\Delta+\omega_{FS}}\right)^2. \label{eq:GammaZZapprox}
\end{equation}

Combining Eqs.~\ref{eq:F0approx} and \ref{eq:GammaZZapprox}, we find
\begin{equation}
    \zeta_{L, LS} \propto \frac{F_0/(\hbar\partial k)}{\Gamma} \approx \frac{\Delta_z}{\gamma}\left(\frac{2\Delta^2 + 2\Delta \omega_{FS} +\omega_{FS}^2}{\omega_{FS}^2}\right).
\end{equation}

Considering detunings $\Delta \gg \omega_{FS}$, which is experimentally reasonable for the smaller fine structure splittings of $^9$Be$^+$ and $^{24}$Mg$^+$, the LS gate figure of merit simplifies to
\begin{equation}
    \zeta_{L,LS}\sim \left(\frac{\Delta_z}{\omega_{FS}}\right)\left(\frac{\Delta^2}{\gamma\omega_{FS}}\right).\label{eq:FOMZZapprox}
\end{equation}
Therefore we see that $\zeta_{L, LS}$ is proportional to the magnetic field and the detuning squared and inversely proportional to the excited state decay rate and the fine structure splitting squared. This result agrees with Figs.~\hyperref[fig:FOMZZlinmultiB]{\ref*{fig:FOMZZlinmultiB}(c)} and \hyperref[fig:MgPlotsLS]{\ref*{fig:MgPlotsLS}(b)}.

\subsection{M\o lmer-S\o rensen gate}
We assume perpendicularly polarized beams of equal intensities, with the higher frequency beam having $\sigma$ polarization and the lower frequency beam having $\pi$ polarization (see configuration 1 in Table~\ref{tab:MS_config}). We will use an approach similar to the above for the LS gate. It is convenient to define the ratios $\omega_{FS}'\equiv\omega_{FS}/\Delta_{MS}$ and $\Delta_z'\equiv\Delta_z/\Delta_{MS}$.

For this configuration, the two-photon Raman Rabi frequency (Eq.~\ref{eq:Omega_R}) is
\begin{align}
    \Omega_R = \frac{g_0^2}{3\Delta_{MS}}\left(\frac{1}{1-\frac{5}{3}\Delta_z'}-\frac{1}{1+\omega_{FS}'-\frac{4}{3}\Delta_z'}\right).\label{eq:OmegaRapprox}
\end{align}
Expanding to first order in $\Delta_z'$ yields
\begin{equation}
    \Omega_R \approx \frac{g_0^2}{3\Delta_{MS}}\left[\frac{\omega_{FS}^\prime}{1+\omega_{FS}^\prime}+\frac{1}{3}\Delta_z^\prime\left(\frac{1+10\omega_{FS}^\prime + 5\omega_{FS}^{\prime 2}}{\left(1+\omega_{FS}^\prime\right)^2}\right)\right].
\end{equation}
\begin{widetext}
We also find the expansions for decoherence due to Raman scattering,

\begin{align}
    \Gamma_{\downarrow\uparrow} &= \frac{g_0^2\gamma}{9\Delta_{MS}^2}\left[2\left(\frac{1}{1-\frac{5}{3}\Delta_z^\prime} -\frac{1}{1 + \omega_{FS}^\prime - \frac{4}{3}\Delta_z^\prime}\right)^2 + \left(\frac{1}{1-\frac{7}{3}\Delta_z^\prime} - \frac{1}{1 + \omega_{FS}^\prime - \frac{8}{3}\Delta_z^\prime}\right)^2 \right]\nonumber\\
    &\approx \frac{g_0^2\gamma}{9\Delta_{MS}^2}\left[3\left(\frac{\omega_{FS}^\prime}{1+\omega_{FS}^\prime}\right)^2 + \frac{2}{3}\left(\frac{\omega_{FS}^\prime}{1+\omega_{FS}^\prime}\right)\Delta_z^\prime\left(\frac{1+34\omega_{FS}^\prime + 17\omega_{FS}^{\prime 2}}{\left(1+\omega_{FS}^\prime\right)^2}\right)\right],\\
    \Gamma_{\uparrow\downarrow} &= \frac{g_0^2\gamma}{9\Delta_{MS}^2}\left[\left(\frac{1}{1-\frac{5}{3}\Delta_z^\prime} - \frac{1}{1 + \omega_{FS}^\prime - \frac{4}{3}\Delta_z^\prime}\right)^2 + 2\left(\frac{1}{1+\frac{5}{3}\Delta_z^\prime} - \frac{1}{1 + \omega_{FS}^\prime +\frac{4}{3}\Delta_z^\prime}\right)^2\right]\nonumber\\
    &\approx \frac{g_0^2\gamma}{9\Delta_{MS}^2}\left[3\left(\frac{\omega_{FS}^\prime}{1+\omega_{FS}^\prime}\right)^2 - \frac{2}{3}\left(\frac{\omega_{FS}^\prime}{1+\omega_{FS}^\prime}\right)\Delta_z^\prime\left(\frac{1 + 10\omega_{FS}^\prime + 5\omega_{FS}^{\prime 2}}{\left(1+\omega_{FS}^\prime\right)^2}\right)\right],\\
    \Rightarrow \Gamma_r &\approx \frac{2g_0^2\gamma}{9\Delta_{MS}^2}\left(\frac{\omega_{FS}^\prime}{1+\omega_{FS}^\prime}\right)^2\left[3 + 4\Delta_z^\prime\left(\frac{2+\omega_{FS}^\prime}{1 + \omega_{FS}^\prime}\right)\right].\label{eq:GammaRXX}
\end{align}

The expression for $\Gamma_{el}$ in terms of $\omega_{FS}^\prime$ and $\Delta_z^\prime$ is
\begin{align}
    \Gamma_{el} &= \frac{g_0^2\gamma}{9\Delta_{MS}^2}\Bigg\{\frac{1}{2}\left[\frac{1}{3}\left(\frac{1}{1-\frac{5}{3}\Delta_z^\prime} + \frac{2}{1 + \omega_{FS}^\prime - \frac{4}{3}\Delta_z^\prime}\right) - \frac{1}{1-\Delta_z^\prime}\right]^2 + \frac{1}{2}\left[\frac{1}{1+\Delta_z^\prime} - \frac{1}{3}\left(\frac{2}{1+\omega_{FS}^\prime + \frac{4}{3}\Delta_z^\prime} + \frac{1}{1+\frac{5}{3}\Delta_z^\prime}\right)\right]^2\nonumber\\
    & + \frac{1}{9}\left[\frac{1}{1+\omega_{FS}^\prime - \frac{8}{3}\Delta_z^\prime} + \frac{2}{1 -\frac{7}{3}\Delta_z^\prime} - \left(\frac{1}{1+\omega_{FS}^\prime - \frac{4}{3}\Delta_z^\prime} + \frac{2}{1 - \frac{5}{3}\Delta_z^\prime}\right)\right]^2\Bigg\}\nonumber\\
    &\approx \frac{4g_0^2\gamma}{81\Delta_{MS}^2}\left(\frac{\omega_{FS}^\prime}{1+\omega_{FS}^\prime}\right)^2.\label{eq:GammaElXX}
\end{align}

Combining $\Gamma_r$ (Eq.~\ref{eq:GammaRXX}) and this result for $\Gamma_{el}$ we calculate the total decoherence rate
\begin{align}
    \Gamma = \frac{1}{4}\left(\Gamma_{el} + 3\Gamma_r\right) \approx \frac{g_0^2\gamma}{18\Delta_{MS}^2}\left(\frac{\omega_{FS}^\prime}{1+\omega_{FS}^\prime}\right)^2\left[\frac{83}{9}+12\Delta_z^\prime\left(\frac{2+\omega_{FS}^\prime}{1+\omega_{FS}^\prime}\right)\right].
\end{align}

Finally, we calculate the figure of merit by dividing the approximate expression for $\Omega_R$ (Eq.~\ref{eq:OmegaRapprox}) by this result for the total decoherence rate and again expanding to first order in $\Delta_z^\prime$. The result, now in terms of $\omega_{FS}$ and $\Delta_z$ rather than $\omega_{FS}^\prime$ and $\Delta_z^\prime$, is
\begin{align}
    \zeta_{L,MS}\approx \frac{54\Delta_{MS}}{83\gamma}\left(\frac{\Delta_{MS} + \omega_{FS}}{\omega_{FS}}\right)^2\left[\frac{\omega_{FS}}{\Delta_{MS} + \omega_{FS}} + \frac{1}{249}\frac{\Delta_z\Delta_{MS}}{\left(\Delta_{MS} + \omega_{FS}\right)^2}\left(83 + 182 \frac{\omega_{FS}}{\Delta_{MS}} + 91\left(\frac{\omega_{FS}}{\Delta_{MS}}\right)^2\right)\right].\label{eq:zetaLMSApprox}
\end{align}
This expression can better be understood when we impose the constraint $\omega_{FS}\ll\Delta_{MS}$, which is attainable for $^9$Be$^+$ and $^{24}$Mg$^+$ and becomes harder to achieve for heavier ions. It is also helpful to note that in general $\omega_{FS}$ will be significantly larger than $\Delta_z$, even for $^9$Be$^+$ ($\omega_{FS}/\Delta_z \sim 3$). With these assumptions, $\zeta_{L,MS}$ simplifies to 
\begin{equation}
    \zeta_{L,MS} \sim \frac{\Delta_{MS}^2}{\gamma \omega_{FS}}.
\end{equation}
We note that, unlike for the LS gate configuration we have considered, the linear figure of merit decreases only linearly with the fine structure splitting, making it a better choice in this configuration for heavier ions. This result agrees well with the numerical results shown in Figs.~\hyperref[fig:MSmultiB]{\ref*{fig:MSmultiB}(b)} and \hyperref[fig:MgPlotsMS]{\ref*{fig:MgPlotsMS}(b)}.

We can also consider the case where $\omega_{FS} \sim \Delta_{MS}$, which would be a more likely operating configuration for heavier ions. For this case, it is straightforward to show that the approximate figure of merit is linear in either $\omega_{FS}$ or $\Delta_{MS}$ (since they are constrained to be similar) and independent of the magnetic field.
\end{widetext}

\section{Exact expressions for the fidelity of a two-qubit entangling gate with decoherence}\label{appendix:fidelity}
Here we provide exact expressions for the two-qubit entangling gate fidelity in the presence of decoherence, for both the light-shift and M{\o}lmer-S{\o}renson implementations.

\subsection{Light-shift gate}
For the light-shift gate, we compute the fidelity between the ideal time-evolved state in Eq.~\eqref{eq:IdealBellZZ}, and state evolved under the light-shift gate, with time evolution described by the master equation
\begin{align}
    \frac{d\hat{\rho}}{dt} = -\frac{i}{2}\left[J\hat{\sigma}_1^z\hat{\sigma}_2^z, \hat{\rho}\right] + \sum_j\mathcal{L}\left[\hat{O}_j\right]\hat{\rho},
    \label{eq:MasterEq_zz}
\end{align}
where $J = J_{12}$ is the coupling strength between the two ions, and $\mathcal{L}\left[\hat{O}_j\right]\hat{\rho}$ is given by Eq.~\eqref{eq:Lindbladian}. We consider jump operators
\begin{align}
    \hat{O}_{\uparrow\downarrow,1} &= \sqrt{\Gamma_{\uparrow\downarrow}}\hat{\sigma}^-_1,\quad \hat{O}_{\uparrow\downarrow,2} = \sqrt{\Gamma_{\uparrow\downarrow}}\hat{\sigma}^-_2, \nonumber\\
    \hat{O}_{\downarrow\uparrow,1} &= \sqrt{\Gamma_{\downarrow\uparrow}}\hat{\sigma}^+_1,\quad \hat{O}_{\downarrow\uparrow,2} = \sqrt{\Gamma_{\downarrow\uparrow}}\hat{\sigma}^+_2, \nonumber\\
    \hat{O}_{el,1} &= \sqrt{\frac{\Gamma_{el}}{4}} \hat{\sigma}^z_1,\quad \hat{O}_{el,2} = \sqrt{\frac{\Gamma_{el}}{4}} \hat{\sigma}^z_2.
    \label{eq:LS_jump_2part}
\end{align}
Now, the ideally evolved density matrix, $\hat{\rho}_{0}(t) = \ket{\psi_0(t)}\bra{\psi_0(t)}$, can be expressed as 
\begin{align}
\begin{split}
    \hat{\rho}_{0}(t) = \frac{1}{4}\Big[1 + \hat{\sigma}_1^x\hat{\sigma}_2^x + \sin(J t)\left( \hat{\sigma}_1^y\hat{\sigma}_2^z + \hat{\sigma}_1^z\hat{\sigma}_2^y\right) \\
    + \cos(J t) \left(\hat{\sigma}_1^x + \hat{\sigma}_2^x\right) \Big].
    \end{split}
\end{align}
Since this state remains pure during the evolution, then the fidelity of the state $\hat{\rho}(t)$ evolved under the light-shift Hamiltonian can be written as
\begin{align}
\begin{split}
    \mathcal{F}_{LS}(t) &\equiv \left(\mathrm{Tr}\sqrt{\sqrt{\hat{\rho}(t)}\hat{\rho}_0(t)\sqrt{\hat{\rho}(t)}}\right)^2
    \end{split}
    \\
    \begin{split}
    &= \mathrm{Tr}\left[\hat{\rho}_{0}(t)\hat{\rho}(t)\right]
    \end{split}
    \\
    \begin{split}
    &= \frac{1}{4} \Bigg[ 1 + \braket{\hat{\sigma}_1^x(t)\hat{\sigma}_2^x(t)}\\
    &\qquad+ \sin(J t)\Big( \braket{\hat{\sigma}_1^z(t)\hat{\sigma}_2^y(t)} + \braket{\hat{\sigma}_1^y(t)\hat{\sigma}_2^z(t)} \Big) \\
    &\qquad+ \cos(J t)\Big( \braket{\hat{\sigma}_1^x(t)} + \braket{\hat{\sigma}_2^x(t)} \Big) \Bigg],
    \end{split}
\end{align}
where expectation values are taken with respect to $\hat{\rho}(t)$, using the fact that $\mathrm{Tr}[\hat{\sigma}_i^\alpha\hat{\sigma}_j^\beta] = \delta_{ij}\delta^{\alpha\beta}$.

To solve the master equation for the required set of correlators, we have the system of equations
\begin{align}
\begin{split}
    \frac{d}{dt}\begin{pmatrix} \braket{\hat{\sigma}_1^+} \\
    \braket{\hat{\sigma}_1^+\hat{\sigma}_2^z} \end{pmatrix}
    = \begin{pmatrix} -\Gamma & iJ \\ 
    iJ - 2\Gamma_- & -(2\Gamma_{+} + \Gamma) \end{pmatrix}\begin{pmatrix} \braket{\hat{\sigma}_1^+} \\ \braket{\hat{\sigma}_1^+\hat{\sigma}_2^z} \end{pmatrix},
    \end{split}
\end{align}
where we have defined $\Gamma_{\pm} = (\Gamma_{\uparrow\downarrow} \pm \Gamma_{\downarrow\uparrow})/2$ and $\Gamma = \Gamma_+ + \Gamma_{el}/2$ for convenience.

For our initial state, this leads to
\begin{gather}
    \braket{\hat{\sigma}^+_1(t)} = \frac{e^{-(\Gamma + \Gamma_+) t}}{2}\Big[\cos\left(t\tilde{J} \right) + \Gamma_+ t \,\,\mathrm{sinc}\left(t\tilde{J} \right)\Big],\\
    \braket{\hat{\sigma}^+_1(t)\hat{\sigma}_2^z(t)} = e^{-(\Gamma + \Gamma_+) t}\left(iJ/2 - \Gamma_{-}\right) t \,\,\mathrm{sinc}\left(t\tilde{J} \right),
\end{gather}
for complex frequency $\tilde{J} = \sqrt{J^2 + 2iJ\Gamma_{-} - \Gamma_{+}^2}$. We note that the Raman scattering term leads to a shift in the effective oscillation rate of these observables. In addition, we find that $\braket{\hat{\sigma}_1^x\hat{\sigma}_2^x}(t) = e^{-2\Gamma t}\braket{\hat{\sigma}_1^x\hat{\sigma}_2^x}(0)$, so we have the result
\begin{widetext}
\begin{align}
\begin{split}
     \mathcal{F}_{LS}(t) = \frac{1}{4}\big(1 + e^{-2\Gamma t}\big) + \frac{e^{-(\Gamma + \Gamma_+) t}}{2}\Bigg[\cos(J t) \mathrm{Re}\Big\{\cos(\tilde{J} t) + \Gamma_+ t\,\,\mathrm{sinc}(\tilde{J} t)\Big\} + 2\sin(J t) \mathrm{Im}\Big\{(iJ/2 - \Gamma_-)t \,\,\mathrm{sinc}(\tilde{J} t)\Big\}\Bigg].\label{eq:F_ls}
     \end{split}
\end{align}
\end{widetext}
To better elucidate the temporal dependence in this expression, we rewrite this as
\begin{widetext}
\begin{align}
\begin{split}
    \mathcal{F}_{LS}(t) = \frac{e^{-(\Gamma + \Gamma_+ +|\tilde{J}|\sin\phi) t}}{8|\tilde{J}|}\Big[\big(|\tilde{J}| - a_+\big)\cos(\tilde{J}_+ t) + \big(|\tilde{J}| - a_-\big)\cos(\tilde{J}_{-}t) + b_+\sin(\tilde{J}_{+}t) + b_-\sin(\tilde{J}_{-}t)\Big] \\
    +  \frac{e^{-(\Gamma + \Gamma_+ - |\tilde{J}|\sin\phi) t}}{8|\tilde{J}|}\Big[\big(|\tilde{J}| + a_-\big)\cos(\tilde{J}_{+}t) + \big(|\tilde{J}| + a_+\big)\cos(\tilde{J}_{-}t) + b_-\sin(\tilde{J}_{+}t) + b_+\sin(\tilde{J}_{-}t)\Big] + \frac{1}{4}\big(1 + e^{-2\Gamma t}\big),
    \end{split}\label{eq:fid_LS}
\end{align}
\end{widetext}
where $\phi = \mathrm{arg}(\tilde{J})$. We have also introduced the real frequencies $\tilde{J}_{\pm} = |\tilde{J}|\cos\phi \pm J$, and defined time-independent coefficients
\begin{gather}
    a_{\pm} = \gamma_{\pm}\sin\phi \pm J\cos\phi,\\
    b_{\pm} = \gamma_{\pm}\cos\phi \mp J\sin\phi,
\end{gather}
with $\gamma_{\pm} = \Gamma_+ \pm 2\Gamma_-$.
In the case where $\Gamma_{\uparrow\downarrow} = \Gamma_{\downarrow\uparrow}$, then $\tilde{J} = J\sqrt{1-(\Gamma_+/J)^2}$ and $\phi = 0$, so the fidelity takes the simpler form
\begin{widetext}
\begin{align}
    \mathcal{F}_{LS}(t) = \frac{1}{4}\Bigg\{ 1 + e^{-2\Gamma t} + \frac{e^{-(\Gamma + \Gamma_+) t}}{\tilde{J}}\Big[\tilde{J}_-\cos\big(\tilde{J}_+ t\big) + \tilde{J}_+\cos\big(\tilde{J}_- t\big) + \Gamma_+\sin\big(\tilde{J}_+ t\big) + \Gamma_+\sin\big(\tilde{J}_- t\big)\Big]\Bigg\}.
\end{align}
\end{widetext}
In both the case of equal or nonequal absorption and decay, we observe that $\mathcal{F}_{LS}(t)$ possesses a contribution decaying with rate $2\Gamma$, as well as an oscillatory term with decay envelope described by decay rate $\Gamma + \Gamma_+$. In the small error rate limit, expanding Eq.~\eqref{eq:fid_LS} yields the linear decay rate of the fidelity $\Gamma_{LS}^{\textrm{fid}} = \Gamma_+ + \Gamma_{el}/2$. This expression matches the denominator used in the linear and quadratic figures of merit for the LS gate. At the Bell state time $\tau_g = \pi/2J$, expanding Eq.~\eqref{eq:fid_LS} yields
\begin{align}
    \mathcal{F}_{LS}(\tau_g) \approx 1- \frac{\tau_g}{2}(3\Gamma_+ + \Gamma_{el}).
\end{align}

\subsection{M{\o}lmer-S{\o}renson gate}
We now consider the case of the M{\o}lmer-S{\o}renson gate. In this case, the ideal time-evolved state may be written as
\begin{align}
\begin{split}
    \hat{\rho}_{0}(t) = \frac{1}{4}\Big[1 + \hat{\sigma}_1^z\hat{\sigma}_2^z + \sin(J t)\left( \hat{\sigma}_1^x\hat{\sigma}_2^y + \hat{\sigma}_1^y\hat{\sigma}_2^x\right) \\
    - \cos(J t) \left(\hat{\sigma}_1^z + \hat{\sigma}_2^z\right) \Big],
    \end{split}
\end{align}
so the fidelity is given by
\begin{align}
\begin{split}
\mathcal{F}_{MS}(t) &= \frac{1}{4} \Bigg[ 1 + \braket{\hat{\sigma}_1^z(t)\hat{\sigma}_2^z(t)} \\
&+ 2i\sin(J t)\Big( \braket{\hat{\sigma}_1^+(t)\hat{\sigma}_2^+(t)} - \braket{\hat{\sigma}_1^-(t)\hat{\sigma}_2^-(t)} \Big)\\
&+ \cos(J t)\Big( \braket{\hat{\sigma}_1^z(t)} + \braket{\hat{\sigma}_2^z(t)} \Big) \Bigg].
\end{split}
\end{align}
We emphasize that we are considering an initial state polarized along $-z$. One could alternatively consider a state polarized along $\pm y$, for which the decoherence would have a different effect on the state evolution, but we do not treat this case here.

To solve for the required set of correlators, we have the following set of equations of motion:
\begin{widetext}
\begin{align}
    \frac{d}{dt}\begin{pmatrix} \braket{\hat{\sigma}_1^z + \hat{\sigma}_2^z} \\ 
    \braket{\hat{\sigma}_1^x\hat{\sigma}_2^y} + \braket{\hat{\sigma}_1^y\hat{\sigma}_2^x} \\
    \braket{\hat{\sigma}_1^z\hat{\sigma}_2^z} \end{pmatrix} = \begin{pmatrix} -2\Gamma_{+} & J & 0 \\ 
    -J & -2\Gamma & 0 \\
    -2\Gamma_{-} & 0 & -4\Gamma_{+} \end{pmatrix}\begin{pmatrix} \braket{\hat{\sigma}_1^z + \hat{\sigma}_2^z} \\ 
    \braket{\hat{\sigma}_1^x\hat{\sigma}_2^y} + \braket{\hat{\sigma}_1^y\hat{\sigma}_2^x} \\
    \braket{\hat{\sigma}_1^z\hat{\sigma}_2^z} \end{pmatrix} - \begin{pmatrix} 4\Gamma_{-} \\ 
    0 \\
   0 \end{pmatrix}.
\end{align}
\end{widetext}
Diagonalization of this linear system and judicious rearrangement and substitution of the resulting expressions leads to the following general result:
\begin{widetext}
\begin{align}
\begin{split}
    \mathcal{F}_{MS}(t) = \frac{1}{4}\left[1 + \frac{\Gamma_-^2\Lambda_+}{\Gamma_+\eta_+}\right] + \frac{e^{-4\Gamma_+t}}{4}\left[1 - \frac{\Gamma_-(2\Gamma_+ - \Gamma_-)\Lambda_-}{\Gamma_+\eta_-}\right] +\frac{\Gamma_-}{2\eta_+}\Big[\Lambda_+\cos(J t) + J\sin(J t)\Big]\\
    -\frac{e^{-(\Gamma + \Gamma_+) t}}{8}\Big[A_+\cos(J_+t) - A_-\cos(J_-t) + B_+\sin(J_+t) - B_-\sin(J_-t)\Big] - \frac{\Gamma_-e^{-(\Gamma + \Gamma_+) t}}{4J^\prime}\Big[C\cos(J^\prime t) + D\sin(J^\prime t)\Big]
    \end{split} \label{eq:fid_MS}
\end{align}
\end{widetext}
where we have grouped terms by their time dependence. We have defined the effective coupling $J' = J\sqrt{1 - (\Gamma_{el}/(2J))^2}$, and introduced the frequency sum/differences $J_{\pm} = J \pm J^\prime$. We have also introduced the parameters $\eta_{\pm} = J^2/2 + 2\Gamma_+^2 \pm \Gamma_+\Gamma_{el}$ and $\Lambda_{\pm} = 2\Gamma_+ \pm \Gamma_{el}$ for compactness of notation. We also have the time-independent, dimensionless coefficients
\begin{gather}
    A_\pm = \frac{1}{\eta_+J^\prime} \Big[J_{\mp}(2\eta_+ - 2\Gamma_-\Gamma_+ - \Gamma_-\Lambda_+) \pm \Gamma_-\Gamma_{el}J^\prime\Big],\\
    B_\pm = \frac{1}{\eta_+ J^\prime}\Big[-2\Gamma_- J J_{\mp} - \Gamma_{el}\eta_+ + \Gamma_{el}\Gamma_-\Lambda_+\Big],\\
    C = \frac{J^\prime}{\eta_+\eta_-}\Big[2\Gamma_{\downarrow\uparrow}(\Gamma_{el}^2-4\Gamma_+^2 - J^2) + J^2\Gamma_{el}\Big],\\
    \begin{split}
    D = \frac{1}{2\eta_+\eta_-}\Big[2\Gamma_{\downarrow\uparrow}\Gamma_{el}(\Gamma_{el}^2 - 4\Gamma_+^2 - 3J^2) \\
    + J^2(\Gamma_{el}^2 - 8\Gamma_+^2 - 2J^2)\Big].
    \end{split}
\end{gather}
In the case that $\Gamma_{\uparrow\downarrow} = \Gamma_{\downarrow\uparrow}$, so that $\Gamma_- = 0$, this expression simplifies to
\begin{widetext}
\begin{align}
\begin{split}
    \mathcal{F}_{MS}(t) = \frac{1}{4}\Big[1+e^{-4\Gamma_+t}\Big] + \frac{e^{-(\Gamma + \Gamma_+) t}}{4J^\prime}\Bigg\{2J^\prime\cos(J_- t) + \sin(J^\prime t)\Big[J_-\sin(J t) + \Gamma_{el}\cos(J t)\Big]\Bigg\}.
    \end{split}
\end{align}
\end{widetext}

In the small error rate limit, the linear decay rate of Eq.~\eqref{eq:fid_MS} is given by $\Gamma_{MS}^{\textrm{fid}} = 2(\Gamma_+ - \Gamma_-) = 2\Gamma_{\downarrow\uparrow}$. While this gate is more susceptible to the presence of Raman decoherence compared to the light-shift gate, and is also far more dependent on the relative imbalance of absorption and decay processes, it is also significantly more robust to the presence of elastic Rayleigh scattering. At the Bell state time $\tau_g = \pi/2J$, expanding Eq.~\eqref{eq:fid_MS} yields
\begin{align}
    \mathcal{F}_{MS}(\tau_g) \approx 1- \tau_g\left[2\Gamma_+ + \Gamma_{el}/4 - (4/\pi)\Gamma_{-}\right].
\end{align}

\subsection{Gate fidelity comparison}
\begin{figure}[t]
    \centering
    \includegraphics[width=\columnwidth]{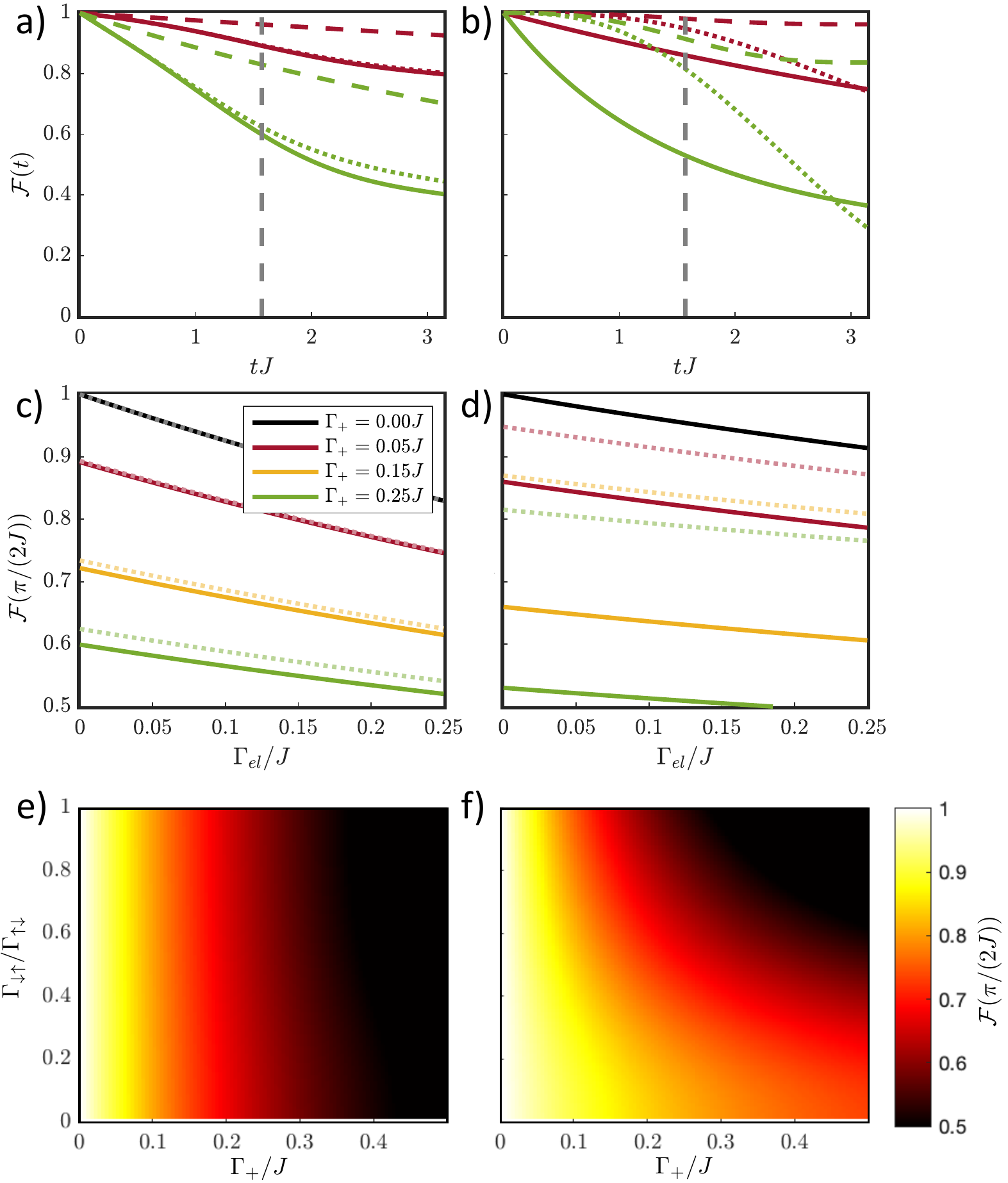}
    \caption{Comparison of the two-qubit gate fidelity for the light-shift gate (left) and M{\o}lmer-S{\o}renson gate (right). (a,~b) Sample dynamics for $\Gamma_+ = 0.05J$ (red), $0.25J$ (green), with solid and dotted lines corresponding to the case of $\Gamma_- = 0$ and $\Gamma_- = \Gamma_+$, respectively ($\Gamma_{el} = 0$). Dashed lines correspond to $\Gamma_{el} = 0.05J$ (red), $0.25J$ (green), with $\Gamma_+ = 0$. Vertical lines denote Bell-state time $tJ = \pi/2$. (c,~d) Fidelity at the Bell-state time for a range of $\Gamma_{el}$ and various values of $\Gamma_+$, with solid and dotted lines corresponding to $\Gamma_- = 0$ and $\Gamma_- = \Gamma_+$, respectively. (e,~f) Fidelity at the Bell-state time for a range of $\Gamma_+$ and $\Gamma_{\downarrow\uparrow}/\Gamma_{\uparrow\downarrow}$, with $\Gamma_{el} = 0$.}
    \label{fig:gatefidelity}
\end{figure}

We briefly compare the expressions Eq.~\eqref{eq:fid_LS} and Eq.~\eqref{eq:fid_MS} for a range of parameters in Figs.~\hyperref[fig:gatefidelity]{\ref*{fig:gatefidelity}(a)} and \hyperref[fig:gatefidelity]{\ref*{fig:gatefidelity}(b)}, showing sample dynamics of the fidelity for various values of $\Gamma_{\uparrow\downarrow}$, $\Gamma_{\downarrow\uparrow}$, and $\Gamma_{el}$. We observe the short-time behavior follows from the expected short-time expansions of the fidelity expressions, with the light-shift gate demonstrating a larger initial decay in the presence of Raman scattering vs Rayleigh scattering, and the M{\o}lmer-S{\o}renson gate displaying an initial linear decay of the fidelity only when $\Gamma_{\downarrow\uparrow} \neq 0$. In Figs.~\hyperref[fig:gatefidelity]{\ref*{fig:gatefidelity}(c)} and \hyperref[fig:gatefidelity]{\ref*{fig:gatefidelity}(d)}, we compare the gate performance for a range of $\Gamma_{el}$, for select values of $\Gamma_{+}$ and $\Gamma_-$, plotting the fidelity at the Bell state time $tJ = \pi/2$. We find the light-shift gate fidelity to be comparatively more susceptible to the presence of elastic scattering than the M{\o}lmer-S{\o}renson gate, while also performing marginally better in the presence of inelastic scattering when $\Gamma_- = 0$. However, we also observe that the fidelity of the M{\o}lmer-S{\o}renson gate is drastically improved when $\Gamma_- = \Gamma_+$, corresponding to the case of $\Gamma_{\downarrow\uparrow} = 0$. We also observe this trend in a systematic comparison of the fidelity at the Bell-state time for a range of $\Gamma_{\downarrow\uparrow}/\Gamma_{\uparrow\downarrow}$ (Figs.~\hyperref[fig:gatefidelity]{\ref*{fig:gatefidelity}(e)} and \hyperref[fig:gatefidelity]{\ref*{fig:gatefidelity}(f)}), where the light-shift gate fidelity appears relatively insensitive to the imbalance of absorption/decay processes, while the performance of the M{\o}lmer-S{\o}renson gate can be drastically altered by the value of this quantity.

\bibliography{References}% Produces the bibliography via BibTeX.

\end{document}